\documentclass[prl,  10pt, aps, twocolumn, superscriptaddress]{revtex4-2}

\usepackage{amsmath,amsfonts,amssymb}

\usepackage[colorlinks=true,breaklinks=true]{hyperref}
\hypersetup{allcolors=[rgb]{0.05 0.05 0.75},linkcolor=[rgb]{0.75 0.05 0.05}}
\usepackage{float}
\usepackage{graphicx}
\usepackage{mathrsfs}
\usepackage{xcolor}
\usepackage{mathtools}
\usepackage{orcidlink}
\usepackage[utf8]{inputenc}
\usepackage{booktabs}
\usepackage{siunitx}
\usepackage{microtype}
\usepackage[caption=false]{subfig}

\newcommand{\bB}{{\bf B}}

\newcommand{\br}{{\bf r}}
\newcommand{\bE}{{\bf E}}

\newcommand{\bk}{{\bf k}}
\newcommand{\bv}{{\bf v}}

\newcommand{\bq}{{\bf q}}

\long\def\exclude#1{}

\newcommand{\Eav}{E_{\rm av}}

\def\parfrac#1#2{{\left(\frac{#1}{#2}\right) }} 

\def\parfrac#1#2{{\left(\frac{#1}{#2}\right) }}

\begin{document}

\title{Magnetic Turbulence Boosts Supernova Signals of Axion--Photon Conversion}

\author{Damiano F.\ G.\ Fiorillo \orcidlink{0000-0003-4927-9850}}
\affiliation{Istituto Nazionale di Fisica Nucleare (INFN), Sezione di Napoli,
Complesso Universitario di Monte Sant’Angelo, Via Cintia, 80126 Napoli, Italy}
\affiliation{Gran Sasso Science Institute (GSSI), L’Aquila, Italy}

\author{\'Angel Gil Muyor \orcidlink{0000-0003-0205-3010}}
\affiliation{Dipartimento di Fisica e Astronomia, Universit\`a degli Studi di Padova, Via Marzolo 8, 35131~Padova, Italy}
\affiliation{Istituto Nazionale di Fisica Nucleare (INFN), Sezione di Padova, Via Marzolo 8, 35131~Padova, Italy}

\author{Georg G.\ Raffelt
\orcidlink{0000-0002-0199-9560}}
\affiliation{Max-Planck-Institut f\"ur Physik, Boltzmannstra\ss e~8, 85748 Garching, Germany}

\author{Edoardo Vitagliano \orcidlink{0000-0001-7847-1281}}
\affiliation{Dipartimento di Fisica e Astronomia, Universit\`a degli Studi di Padova, Via Marzolo 8, 35131~Padova, Italy}
\affiliation{Istituto Nazionale di Fisica Nucleare (INFN), Sezione di Padova, Via Marzolo 8, 35131~Padova, Italy}

\begin{abstract}
Magnetic fields between a supernova (SN) and Earth convert axions into gamma rays. The absence of such a signal in coincidence with SN~1987A neutrinos, using the coherent Milky Way field, provides well-studied constraints on $g_{ap}\times g_{a\gamma}$ (axion-proton times axion-photon couplings) and on $g_{a\gamma}$ alone. We show that the small-scale power of the turbulent magnetic field component boosts axion–photon conversion and, crucially, extends sensitivity to larger masses. The turbulent field components of the Milky Way and of the Large Magellanic Cloud (hosting SN~1987A) yield improvements of up to 2 orders of magnitude in $g_{ap}\times g_{a\gamma}$. Turbulence should likely impact the sensitivity of other searches based on other axion-photon conversion sites, such as starburst galaxies.
\end{abstract}

\maketitle

\textbf{\textit{Introduction}}---Axions are among the best motivated new particles because they could explain CP conservation in QCD~\cite{Peccei:1977hh, Peccei:1977ur,Weinberg:1977ma, Wilczek:1977pj, Kim:1979if, Shifman:1979if, Dine:1981rt, Zhitnitsky:1980tq, DiLuzio:2020wdo} and also could be the dark matter of the universe~\cite{Preskill:1982cy,Abbott:1982af,Dine:1982ah, OHare:2024nmr}. Other axion-like pseudoscalar bosons are predicted by string theory compactifications, realizing the so-called axiverse~\cite{Svrcek:2006yi,Arvanitaki:2009fg,Cicoli:2012sz,Demirtas:2018akl,Halverson:2019cmy,Demirtas:2021gsq,Gendler:2023kjt}. The smallness of the axion mass, $m_a$, opens search strategies with low-energy techniques and astrophysical methods. The generic axion--two--photon interaction, $g_{a\gamma} a\,\bE\cdot\bB$, enables $a\to\gamma$ conversion in macroscopic $B$-fields \cite{Sikivie:1983ip, Raffelt:1987im}, which is the basis for most systematic axion searches \cite{Graham:2015ouw, Irastorza:2018dyq, Sikivie:2020zpn, Berlin:2024pzi, Baryakhtar:2025jwh} and, through astrophysical $B$-fields, for many constraints~\cite{OHare}, and possible future detection, e.g., by SN $\gamma$ rays \cite{Carenza:2021alz, Manzari:2024jns, Fiorillo:2025gnd, Candon:2025sdm}.

Axions interact extremely feebly, so the $B$-field-induced conversion probability, $P_{a\gamma}=(g_{a\gamma}B L/2)^2$, is strong only if the path length $L$ is large. However, $P_{a\gamma}$ drops quickly when $L\gg k_a^{-1}$, where $k_a=m_a^2/2E_a$ is the axion--photon momentum mismatch. The resulting ``loss of coherence'' in $a$--$\gamma$ conversion bedevils all axion searches: the conversion volume or path length should be large to overcome the smallness of $g_{a\gamma}$ but is limited by $k_a^{-1}$. One way to reach larger masses is to give photons an effective $m_\gamma\simeq m_a$ through a dispersive medium \cite{vanBibber:1988ge}. Alternatively, spatial $B$-field variations with strong Fourier power at $k_a$ would have a similar effect~\cite{VanBibber:1987rq}.

\begin{figure}
    \centering
    \includegraphics[width=\columnwidth]{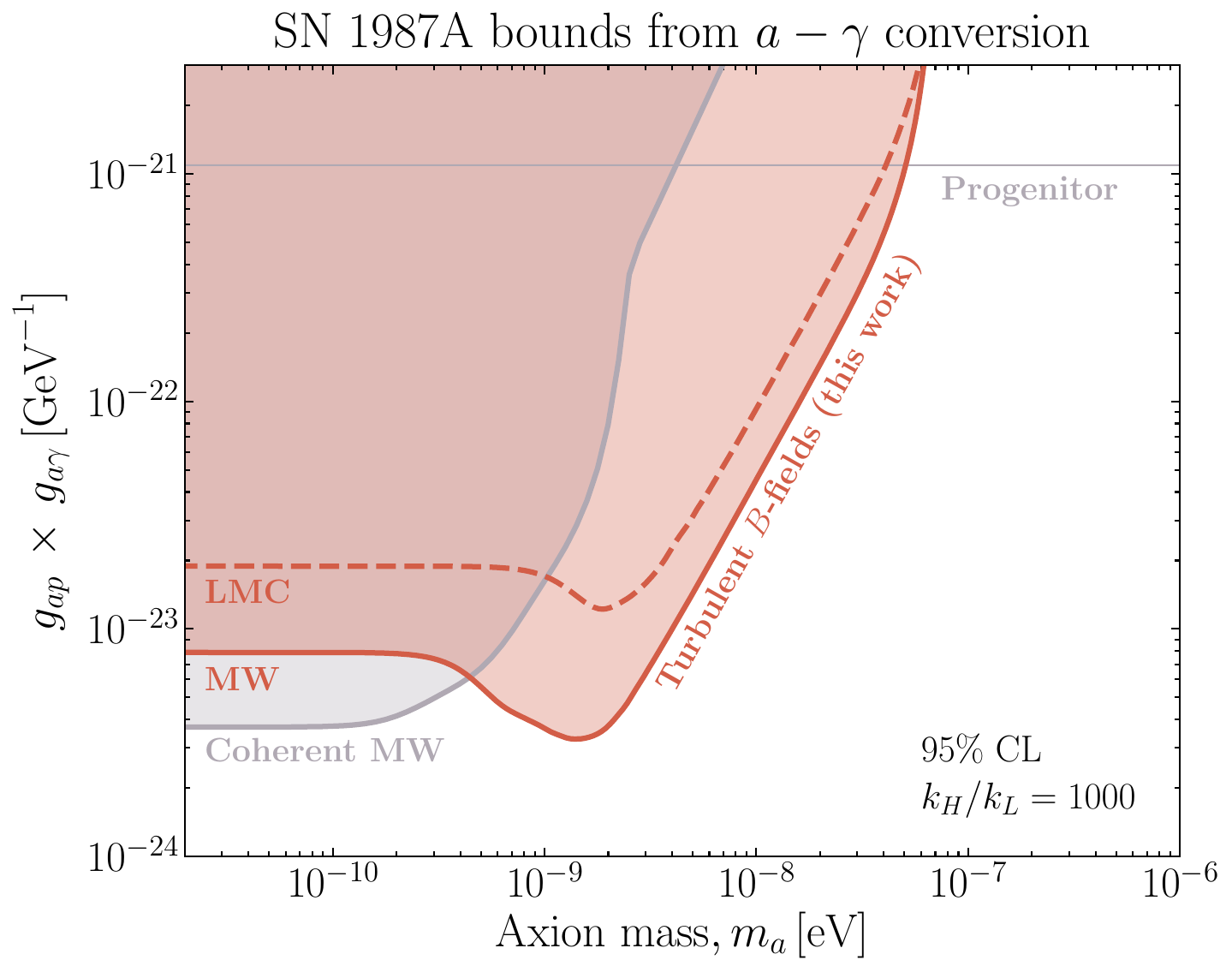}
    \caption{New constraints (red) on $g_{ap}\times g_{a\gamma}$ from non\-detection of SN~1987A gamma rays. We assume $np\to npa$ bremsstrahlung production and $a\to\gamma$ conversion in the turbulent $B$-fields of the MW and LMC with $k_H/k_L=1000$ for the ratio of high and low turbulence scales. In gray, we show previous SN~1987A constraints, labeled ``Coherent MW'' and ``Progenitor'', resulting, respectively, from conversions in the coherent MW field and in the putative dipole field of Sanduleak $-69\,202$, the progenitor of SN~1987A (assumed radius $30\, R_\odot$ and surface magnetic field $30\, \mathrm{G}$)~\cite{Fiorillo:2025gnd}.
    }
    \label{fig:BremssBounds}
\end{figure}

In this Letter, we show that the turbulent component of the Milky Way (MW) $B$-field, in the manner of a magnetic undulator, opens a new $m_a$ range, notably for converting axions from the historical SN~1987A. This peculiar Type~II SN resulted from the collapse of the blue supergiant Sanduleak $-69\,202$ in the Large Magellanic Cloud (LMC) at a distance of $51.4\,\rm kpc$~\cite{Panagia:1998, Panagia:2003} (see Ref.~\cite{2019Natur.567..200P} for a slightly smaller distance). The unique neutrino observations \hbox{\cite{Kamiokande-II:1987idp, Hirata:1988ad, Bionta:1987qt, IMB:1988suc, Alekseev:1987ej, Alekseev:1988gp, Koshiba:1992yb, Fiorillo:2023frv}} provided many particle constraints through the duration of neutrino cooling \cite{Raffelt:1996wa, Raffelt:2025wty, Arza:2026rsl}, notably $g_{aN}\alt 10^{-9}$ for the axion-nucleon coupling \cite{Mayle:1987as, Turner:1987by, Burrows:1988ah, Mayle:1989yx, Burrows:1990pk, Raffelt:2006cw, Carenza:2019pxu, Fiorillo:2025gnd}. In addition, the absence of a contemporaneous $\gamma$-ray signal in the Solar Maximum Mission (SMM) \hbox{satellite \cite{forrest1980gamma, Chupp:1989kx, Kolb:1988pe, Oberauer:1993yr}} provided constraints on $a\to\gamma$ conversion in the MW $B$-field \cite{Brockway:1996yr, Grifols:1996id, Payez:2014xsa, Hoof:2022xbe}. Assuming axion production through their proton coupling in $np\to npa$ bremsstrahlung, the latest bounds on $g_{ap}\times g_{a\gamma}$ \cite{Fiorillo:2025gnd} are shown in Fig.~\ref{fig:BremssBounds}. Conversion in the presumed progenitor field was recently added \cite{Manzari:2024jns, Fiorillo:2025gnd}, extending sensitivity to larger masses due to the smaller spatial scale of $B$-field variation. For self-consistency, we have recomputed these constraints with the axion flux spectrum adopted here (see below).

In a similar vein, we here include another previously ignored component, the turbulent MW field. The impact of $B$-field stochasticity or turbulence on astrophysical \hbox{$a\to\gamma$} conversion was previously
studied in the context of TeV gamma rays~\cite{Mirizzi:2007hr, Meyer:2014epa, Marsh:2021ajy, Carenza:2022zmq}. For SN axions, the relevance of a turbulent MW field was proposed in Ref.~\cite{Carenza:2021alz}, mostly focusing on a possible boost 
(up to a factor of 2) compared to the regular field and the spectral irregularities that would appear in a $\gamma$-ray signal from the next Galactic SN for massive axions.

We show that the gains from including turbulent fields in the conversion go far beyond this effect. For the first time, we derive constraints from SN 1987A including the turbulent fields of the MW and of the LMC, separately. For large axion masses, the bounds on  $g_{ap}\times g_{a\gamma}$ improve by up to 2 orders of magnitude for neV-range masses, as the turbulent magnetic field provides small-scale structure that facilitates resonant $a$--$\gamma$ conversion.

\textbf{\textit{Axion--photon conversion}}---In all cases of interest here, the conversion probability is small, $P_{a\gamma}\ll1$, and back-conversion is irrelevant. In this perturbative regime, the conversion amplitude to a photon with polarization~$j$ is, up to a global phase, the line of sight (LoS) integral
\begin{equation}\label{eq:amplitude}
\mathcal{A}_{a\gamma_j}=\frac{g_{a\gamma}}{2} \int_0^L ds\,B_j(s)\,e^{-i k_a s},
\end{equation}
where $k_a=m_a^2/2E$, the photon plasma mass can be neglected, and $B_j(s)$ are the transverse field components. The conversion probability, summed over both polarizations, then is
\begin{equation}
    P_{a\gamma}(k_a)=\frac{g_{a\gamma}^2}{4}\!\int_0^L\!\!ds \int_0^L\!\!ds'\,
    \bB_\perp(s)\cdot\bB_\perp(s')\,\cos[k_a(s-s')],
    \label{eq:totalprob}
\end{equation}
where $\bB_\perp$ is the field transverse to the LoS. We denote with $P_{a\gamma}(k_a)$ the conversion probability of an axion with momentum mismatch $k_a$, and with $P_{a\gamma}$, implicitly depending on $m_a$, a spectral average. Both quantities coincide for $m_a=0$. Moreover, we use $\langle P_{a\gamma}(k_a)\rangle$ for an average over $B$-field configurations. Axions from SN~1987A traveled through one specific realization of the turbulent $B$-field. The conversion probability $P_{a\gamma}(k_a)$ is one draw from a distribution with average $\langle P_{a\gamma}(k_a)\rangle$.

For massless axions, 
\smash{$P_{a\gamma}=(g_{a\gamma}/2)^2\bigl|\int\! ds\,\bB_\perp(s)\bigr|^2\!$}. For the LoS to SN~1987A, the result \cite{Fiorillo:2025gnd} is given in Table~\ref{tab:Bfields}, derived from the baseline case of the latest coherent MW $B$-field models of Unger and Farrar~\cite{Unger:2023lob}. The SN has Galactic coordinates $b_{\rm SN}=-31.8^\circ$ and $l_{\rm SN}=279.7^\circ$, high above the plane. The $B_j(s)$ components along the LoS are shown in Fig.~6 of Ref.~\cite{Fiorillo:2025gnd}. The transverse parts are somewhat weak relative to the longitudinal one, and sizeable fields occur only for about 20~kpc near the Galactic plane. The same $P_{a\gamma}$ descends from a uniform field with $BL=3.0~\mu{\rm G}~{\rm kpc}$.

For $m_a$ so large that $k_a^{-1}\ll L$, there are  \hbox{many $a$--$\gamma$} oscillations within $L$ so that $\mathcal{A}_{a\gamma}$ mostly cancels. For $L=10~{\rm kpc}$, this occurs for $k_a\gg 6\times10^{-28}~{\rm eV}$, which for $E=100~{\rm MeV}$ implies $m_a\gg 0.4~{\rm neV}$. For a uniform field, one finds $P_{a\gamma}(k_a)=(g_{a\gamma} B)^2 k_a^{-2}\sin^2(k_aL/2)$, which for ${k_a\to0}$ returns to $(g_{a\gamma} B L/2)^2$, while for $k_a\gg L^{-1}$,   $P_{a\gamma}(k_a)$ oscillates around $(g_{a\gamma} B)^2 k_a^{-2}/2$ . For inhomogeneous fields, the same $k_a^{-2}$ scaling is obtained when $k_a^{-1}$ is far below any typical scale of $B$-field variation. 

\textbf{\textit{Turbulent fields}}---The MW has a strong turbulent field (Table~\ref{tab:Bfields}), mostly inferred from the excess of synchrotron radiation on top of the one expected from the regular field measured through Faraday rotation. It was obtained in Ref.~\cite{Jansson:2012rt} and later updated with \textit{Planck} observations~\cite{Planck:2016gdp}, where it was modeled as a Gaussian random field with a Kolmogorov power spectrum and an outer scale equal to $2\pi/k_L=100\, \mathrm{pc}$. This value is consistent with that inferred in the halo from the fluctuations in the Faraday rotation measure~\cite{Mertsch:2013pua,Beck:2014pma}, although it could be lower in the spiral arms~\cite{Haverkorn:2008tb}; this is not relevant for this work due to
SN~1987A high Galactic latitude. We do not include the striated disk field, introduced in Refs.~\cite{Jansson:2012rt,Planck:2016gdp} to boost synchrotron radiation compared to the field implied by Faraday rotation. The striated field may not be needed---its effect can also be mimicked by a correlation between cosmic-ray density and $B$ field~\cite{Jansson:2012rt} or the effect of the Local Bubble~\cite{Korochkin:2024yit,Unger:2025xkk}. In any case, for the high-latitude SN~1987A, it is likely unimportant.

\begin{table}[t]
\vskip-6pt
\caption{Adopted $B$-field parameters and $a\to\gamma$ conversion probability, $P_{a\gamma}$, for massless axions with $g_{a\gamma}=10^{-11}~{\rm GeV}^{-1}$. The turbulent case uses Eq.~\eqref{eq:TurbProb1}. The coherent MW case uses an integration over the actual profile (see text), whereas the stated $|B|$ and $L$ were chosen to provide the same $BL$ and the same high-$k_a$ suppression of a uniform field.
\label{tab:Bfields}}
\begin{tabular*}{\columnwidth}{@{\extracolsep{\fill}} lllll}
\toprule
     &$|B|$, $B_0$&$k_L^{-1}$&$L$& $P_{a\gamma}$,\,$\langle P_{a\gamma}(k_a)\rangle$\\
     &[$\mu$G] & [pc] & [kpc] & \\
\midrule
\multicolumn{1}{l}{Milky Way}\\
\qquad Coherent \cite{Unger:2023lob}                  & 0.15&  ---  &  20  & $2.1\times10^{-3}$  \\
\qquad Turbulent \cite{Jansson:2012rt,Planck:2016gdp} & 5  &  250  &  4  & $3.7\times10^{-3}$  \\
\midrule
\multicolumn{1}{l}{Large Magellanic Cloud}\\
\qquad Coherent \cite{Mao:2012hd, Gaensler:2005qj}      & 1  &  ---  &   1  & $2.3\times10^{-4}$  \\
\qquad Turbulent \cite{2024MNRAS.535.1944L}           & 5  & 200    &   0.8& $5.9\times10^{-4}$  \\
\bottomrule
\end{tabular*}
\end{table}

For $a$--$\gamma$ conversion, a $B$-field undulating with a given wave\-number $k_B$ resonantly enhances $P_{a\gamma}$ for $k_a\simeq k_B$. If the Fourier modes are randomly oriented, in projection on the LoS, the resonance condition is met for any ${0\leq k_a\leq k_B}$, leading to a nearly flat $\langle P_{a\gamma}(k_a)\rangle$ for $k_a\alt k_B$, and to the usual $k_a^{-2}$ variation for $k_a\gg k_B$. The average is taken over field configurations.

The turbulent field is modeled as an isotropic random collection of Fourier modes with a power-law spectrum $|B_k|^2\propto k^{-\beta-2}$ in a range $k_L\lesssim k \lesssim k_H$ and  otherwise zero. The dissipation scale $k_H$ cannot be determined observationally---we show in the Supplemental Material (SupM)~\cite{SM}\nocite{CAST:2024eil,Noordhuis:2022ljw,kolmogorov1991local,padmanabhan2000theoretical,Haverkorn_2013} that $k_H/k_L=1000$ is a conservative analytical estimate, to be used in our analysis,
while observations hint at much larger values \cite{Armstrong:1995zc}. A Kolmogorov spectrum has \hbox{$\beta=5/3$}; Goldreich-Sridhar turbulence~\cite{Goldreich:1994zz} has an analogous spectrum for the wave number perpendicular to the local magnetic field. The value of this index determines the shape of the constraints for axion masses at which conversion is resonantly enhanced. For $k_H^{-1}<k_L^{-1}\ll L$, the $a$--$\gamma$ conversion depends only on the statistical properties of the $B$-field as worked out in the SupM. 

For low-mass axions, one finds a flat function of $k_a$, as anticipated earlier,
\begin{equation}\label{eq:TurbProb1}
    \langle P_{a\gamma}(k_a)\rangle= g_{a\gamma}^2 B_0^2 L\, \frac{\pi(\beta-1)}{8\beta k_L}
    \quad \mathrm{if}\, k_a\ll k_L.
\end{equation}
Here $B_0^2=\langle |\bB(\br)|^2\rangle$, and we use the schematic values for the injection scale $k_L$ collected in Table~\ref{tab:Bfields}; for the LMC, this is extracted from Ref.~\cite{Seta:2022uoy}. For the MW, our choice of $k_L$ is somewhat larger than the typical values in Refs.~\cite{Jansson:2012rt,Planck:2016gdp,Mertsch:2013pua,Beck:2014pma}, which range from $16\,\mathrm{pc}\lesssim k_L^{-1}\lesssim 175\,\mathrm{pc}$.

For larger $k_a$, fewer $B$-field modes are on resonance, and the average conversion efficiency declines with the same power law as the turbulence spectrum,
\begin{align}\label{eq:TurbProb2}
    \langle P_{a\gamma}(k_a)\rangle =g_{a\gamma}^2 B_0^2\,  L\, &\frac{(\beta^2-1)}{4\beta(\beta+2)k_L}\parfrac{k_L}{k_a}^\beta
    \nonumber
    \\[1ex]
   & \kern2em \mathrm{if}~ k_L\ll k_a\ll k_H.
\end{align}
For yet larger $k_a$, the now-familiar $k_a^{-2}$ scaling sets in,
\begin{equation}
    \langle P_{a\gamma}(k_a)\rangle= \frac{g_{a\gamma}^2B_0^2}{3}\,  \frac{1}{k_a^2} 
    \quad \mathrm{if}~ k_H\ll k_a,
    \label{eq:TurbProb3}
\end{equation}
independent of any length scale associated with the $B$-field configuration.

\textbf{\textit{LMC magnetic field}}---The SN~1987A host galaxy also contributes to conversion.  The coherent LMC field was measured with polarized synchrotron emission at 1.4~GHz \cite{Mao:2012hd}. Fermi-LAT $\gamma$-ray detections imply, without assuming equipartition, $B\alt 7\,\mu\mathrm{G}$ after summing the plane-of-the-sky and LoS components, consistent with $ 2\,\mu\mathrm{G}$ expected from equipartition~\cite{Mao:2012hd}. Slightly smaller values of $\sim1\, \mu\mathrm{G}$ derive from Faraday rotation~\cite{Gaensler:2005qj}. A larger field of $\sim 11\, \mu\mathrm{G}$ exists along two filaments, but the LoS to SN~1987A does not intercept these structures. 

The small-scale LMC field is studied in Ref.~\cite{Seta:2022uoy}, where Faraday rotation from 250 different sources is isolated through the structure-function method with higher-order stencils. Assuming isotropic Gaussian random fields and log-normal thermal electron densities, the magnetic energy is found to follow a power law  consistent with Kolmogorov shape. The general properties that we use here are collected in Table~\ref{tab:Bfields}. Finally, the LoS depth to SN~1987A obtained from pulsar dispersion measures is $L\sim 800\, \mathrm{pc}$ for the region RII in Table 5 of Ref.~\cite{2024MNRAS.535.1944L}, where Sanduleak was located. A larger value of several kpc was gathered from red clump star observations~\cite{Subramanian:2008pi}, a value sometimes used in cosmic-ray studies \cite{Reynoso-Cordova:2025meg}. To be conservative, we use the smaller value that we show in Table~\ref{tab:Bfields}.

\textbf{\textit{SN~1987A as an axion source}}---Axions are \hbox{emitted} by the proto--neutron star (PNS) that has formed in the core after its collapse. The main relevant PNS parameters are its internal temperature (which depends on the assumed equation of state), baryon density, total mass, and cooling speed. These properties cannot be inferred from SN~1987A observations and the neutrino signal is also not very constraining \cite{Fiorillo:2023frv}. Among a suite of PNS models~\cite{Fiorillo:2025gnd}, we select their Cold Model, which produces a conservative axion emission. It has an average $T\simeq30~{\rm MeV}$ and final baryonic mass of $1.35\,M_\odot$. 

For production, we assume either Primakoff scattering or nucleon bremsstrahlung. Inspired by the KSVZ model, we assume that axions couple only to protons with a structure $(g_{ap}/2m_p)\,\overline{p}\gamma^\mu \gamma^5 p \,\partial_\mu a$. We ignore pion production, $\pi^-p\to n a$, because of the very uncertain $\pi^-$ abundance \cite{Fiorillo:2025gnd}. We may then use directly the time-integrated spectral fluxes that were provided in the following form~\cite{Fiorillo:2025gnd}:
\begin{equation}\label{eq:GammaSpectrum}
    \frac{dN_a}{dE_a}=N_a\, \frac{(\alpha+1)^{\alpha+1}}{\Gamma(\alpha+1)}\,
    \frac{E_a^\alpha}{\Eav^{\alpha+1}}\,
    \exp\left[-\frac{(1+\alpha)E_a}{\Eav}\right],
\end{equation}
where $N_a$ is the total number of emitted axions, $\Eav$ their average energy,  \hbox{$\alpha=(2-\langle E^2\rangle/\Eav^2)/(\langle E^2\rangle/\Eav^2-1)$} the pinching parameter, and $\Gamma$ the Gamma function (Table~\ref{tab:Fluxes}). Despite a different microphysical treatment, the flux parameters of the Bari group for the Cold Model \cite{Lella:2024hfk} agree reasonably well with ours~\cite{Fiorillo:2025gnd}.

\begin{table}[t]
\vskip-6pt
\caption{Axion flux parameters for Eq.~\eqref{eq:GammaSpectrum}, inspired by the Cold Model of Ref.~\cite{Fiorillo:2025gnd}, with $g=g_{a\gamma}{\rm GeV}$ for Primakoff production and $g=g_{ap}$ for bremsstrahlung. $\mathcal{N}_a$ is the number of axions passing through the SMM effective area of $90~{\rm cm}^2$ in the 25--100~MeV channel. $P^{a\gamma}_{95}$ is the SMM limit (95\% CL) on the $a$--$\gamma$ conversion probability.
\label{tab:Fluxes}}
\vskip2pt
\begin{tabular*}{\columnwidth}{@{\extracolsep{\fill}} llllll}
\toprule
     Source reaction & $N_a$ & $E_{\rm av}$ & $\alpha$&$\mathcal{N}_a$&$g^{2}P^{a\gamma}_{95}$\\[0.4ex]
            & $[10^{73}g^2]$ &[MeV]& &$[10^{27}g^2]$&$[10^{-28}]$\\
\midrule
Primakoff ($\gamma p\to pa$) & 1.6 & 86 & 2.1 & 2.8 & 71 \\
Brems ($np\to npa$)          & 34  & 59 & 1.6 & 69  & 2.9\\
\bottomrule
\end{tabular*}
\end{table}

\textbf{\textit{Fluctuations of the conversion probability---}}The stochastic nature of turbulent fields causes fluctuations of $P_{a\gamma}$, depending on LoS and also on axion energy through $k_a$, as detailed in the SupM. This effect is particularly pronounced for massless axions ($k_a=0$). Assuming Gaussian distributions, $\mathcal{A}_{a\gamma}$ in Eq.~\eqref{eq:amplitude} follows an independent Gaussian for each photon polarization, implying an exponential distribution for $P_{a\gamma}(k_a)|_{k_a=0}$. The energy-averaged version $P_{a\gamma}|_{m_a=0}$ also follows an exponential distribution; therefore, it is affected by large LoS-dependent fluctuations. The physical origin is clear: at all energies, axions probe in essence the power at a single wave number ($k_a=0$), which has large chances of fluctuating.

For nonvanishing $m_a$, implying a nonvanishing and energy-dependent $k_a$, these fluctuations diminish because resonant conversion happens for all wavenumbers spanned by the energy spectrum---it is unlikely that for a given LoS, all modes are simultaneously suppressed. Fluctuations are suppressed when $k_a\gtrsim 2\pi/L$, a regime in which the conversion probability is distributed as $\propto P_{a\gamma}^{\kappa-1} e^{-\kappa P_{a\gamma}/\langle P_{a\gamma}\rangle}$. For large values of the shape parameter $\kappa$, it features a much smaller dispersion compared to the exponential distribution, which itself corresponds to $\kappa=1$. In the heavy-mass regime ($k_a\gtrsim k_H$), there is no resonance, and $P_{a\gamma}$ is once more exponentially distributed. The intermittency of turbulence, which breaks its Gaussianity, would likely increase the chances of large $P_{a\gamma}$ values, but for deriving constraints, we stick to the conservative Gaussian assumption.

\begin{figure}
    \centering
    \includegraphics[width=\linewidth]{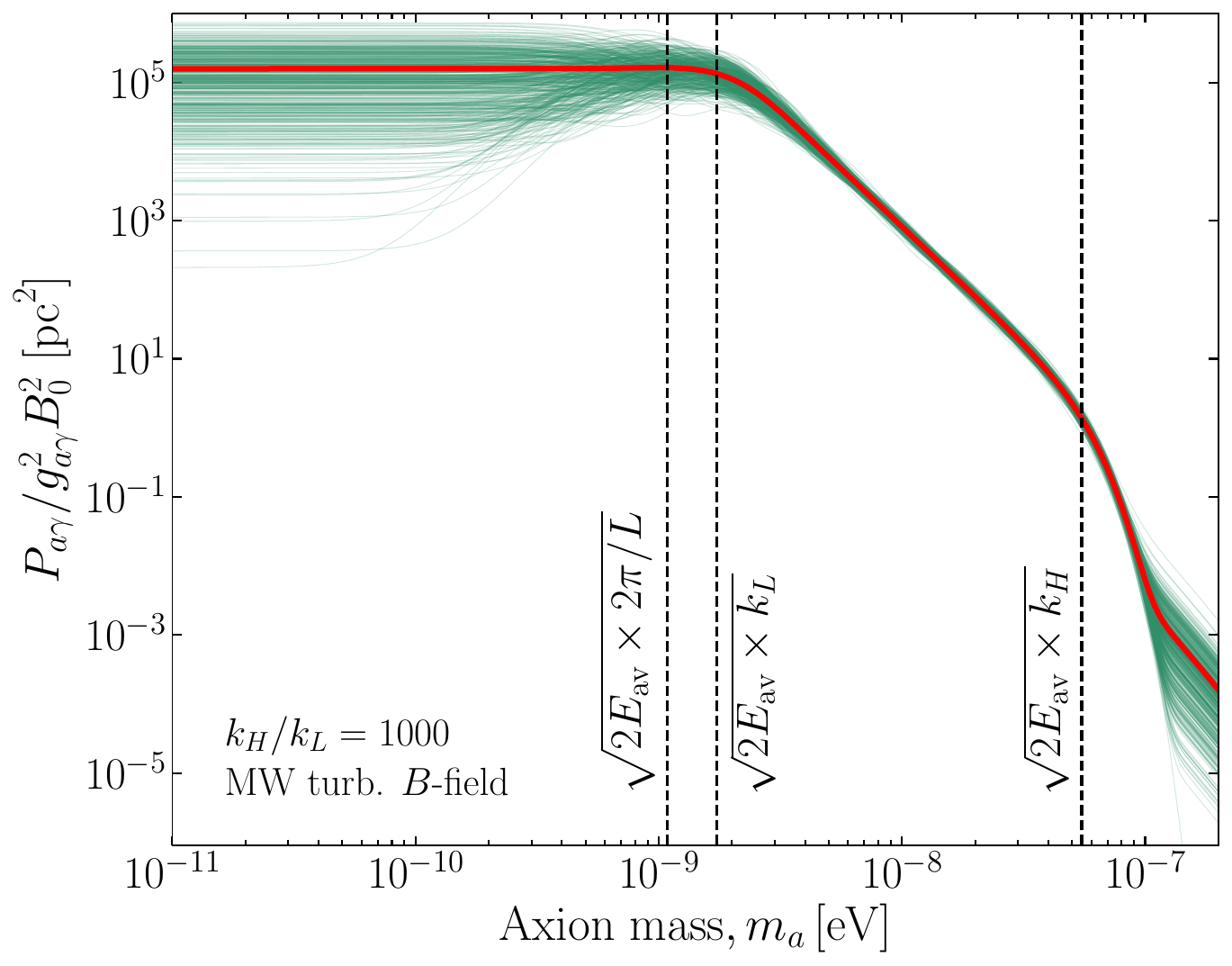}
    \caption{Representative Monte Carlo results for $P_{a\gamma}$, the
    $a$--$\gamma$ conversion probability, for 500 of our $10^4$ realizations, using the bremsstrahlung spectral axion flux. The red line is $\langle P_{a\gamma}\rangle$. Vertical lines indicate the characteristic axion masses at which a typical $k_a$ (evaluated at $E=E_{\rm av}$) is comparable to the inverse of the length, $L$, of the conversion region, the injection scale $k_L$, or the dissipation scale of turbulence~$k_H$.
}
    \label{fig:mc}
\end{figure}

We have validated our estimates with a Monte Carlo simulation of axion conversion in a synthetic turbulent Gaussian field. We have computed $P_{a\gamma}$ from $10^4$ realizations for a grid of $k_a$ values, and convoluted these probabilities with the thermal spectrum of Eq.~\eqref{eq:GammaSpectrum}. The results are shown in Fig.~\ref{fig:mc}. Our simulations confirm the large LoS-dependent fluctutations of $P_{a\gamma}$ for $k_a< 2\pi/L$ and $k_a\gtrsim k_H$. On the other hand, in the intermediate regime, fluctuations are indeed diminished.

\textbf{\textit{Results---}}The SMM satellite did not see any excess $\gamma$ rays coincident with the neutrino burst, tagged by the time of the first event in the IMB detector which had a reliable clock. Following Ref.~\cite{Fiorillo:2025gnd}, we use the 25--100~MeV channel with an effective area of $90~{\rm cm}^2$. In Table~\ref{tab:Fluxes}, we show the number of axions, $\mathcal{N}_a$, passing through this area and with energies in this range. Within three time bins (length 2.048~s each) encompassing the axion signal that lasted for less than 5~s, SMM counted $N_{\rm obs}= 60$~events, with an expected background of $B=54$ \cite{Fiorillo:2025gnd}, consistent with no additional events. Assuming a Poisson distribution for $N_{\rm obs}$, these \hbox{values} imply a constraint on a possible signal $S<20.1$ (95\% CL) \cite{Fiorillo:2025gnd}. One expects $S=P_{a\gamma}\,\mathcal{N}_a$ signal events, leading to the shown limits on~$P_{a\gamma}$. In the massless limit, one finds from the coherent MW field $P_{a\gamma}=2.1\times10^{-3}(g_{a\gamma}{\rm GeV})^2$ (Table~\ref{tab:Bfields}), implying $|g_{ap}\times g_{a\gamma}|<3.7\times10^{-24}\,{\rm GeV}^{-1}$ (95\%~CL). For larger masses, $P_{a\gamma}(k_a)$ depends on energy through $k_a=m_a^2/2E$ so that one needs to convolve the spectrum with $P_{a\gamma}(k_a)$. 

For the turbulent fields, we use the parameters specified in Table~\ref{tab:Bfields} and $k_H/k_L=1000$. We stress that, while we have observational information on $k_L$, the value of $k_H$ and then of the ratio $k_H/k_L$ cannot be measured. As we argue in the SupM, $k_H=1000\, k_L$ is a conservative choice. If this ratio were larger, the resonant enhancement would extend to larger axion masses, as the value of $m_a$ at which the transition between the intermediate regime and the heavy regime occurs scales as $\sim \sqrt{k_H/k_L}$.

In this case, the expected number of events $S$ is itself a random variable, which we model as a Gamma distribution $\Gamma(\kappa,\langle S\rangle)$ and which then has to be convolved with the Poisson distribution. We use our Monte Carlo simulations to extract, for each value of $m_a$, the value of the shape parameter $\kappa$, using the \hbox{25--100~MeV} channel of SMM. Therefore, in this case the expected number of SMM signal events depends also on $m_a$. As an example, for $m_a=3\, \mathrm{neV}$ and in the MW case, we find $\overline S<24.1$ (95\%~CL). Using the value of $\langle P_{a\gamma}\rangle$ given by our simulations, which matches our analytical predictions, this translates into $|g_{ap}\times g_{a\gamma}|<6.4\times10^{-24}\,{\rm GeV}^{-1}$. Doing so for each $m_a$ value leads to the constraints shown in~Fig.~\ref{fig:BremssBounds}. 

The large fluctuations of $\overline S$ as a function of $B$-field configuration weaken the turbulent bounds considerably for small $m_a$. While in the massless limit, the MW turbulent $\langle P_{a\gamma}\rangle$ is larger than  the coherent one, the turbulent constraint is $\overline S<158.8$ (95\%~CL) and the bound on $|g_{ap}\times g_{a\gamma}|$ a factor of 2 weaker. For larger $m_a$, on the other hand, due to the resonant enhancement and the smaller fluctuations in the turbulent case, the turbulent bounds vastly supersede the coherent ones. Moreover, for a larger CL, the bounds will diminish much more for massless axions because of the larger $P_{a\gamma}$ fluctuations. This issue is further discussed in the SupM. For both coherent and turbulent fields, the strongest constraints derive from the MW, as shown in~Fig.~\ref{fig:BremssBounds}.

\begin{figure}
    \centering
    \includegraphics[width=\linewidth]{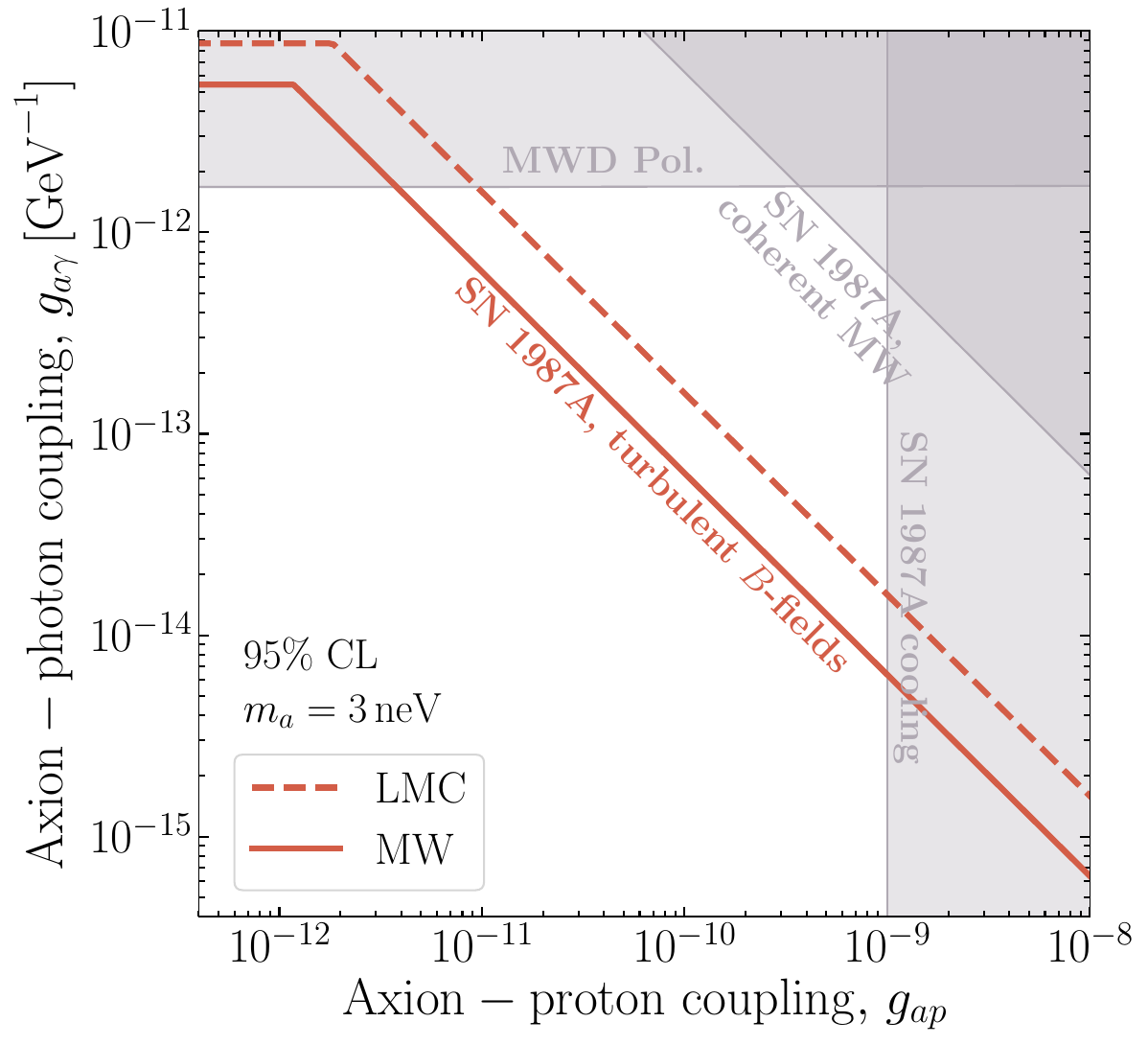}
    \caption{Constraints in the $g_{ap}$--$g_{a\gamma}$ plane for fixed $m_a=3\, \mathrm{ neV}$. Existing limits include the bound from SN~1987A cooling \cite{Raffelt:1996wa, Fiorillo:2025gnd} and from the nonobservation of axion-induced linearly polarized light from magnetic white dwarfs~\cite{Benabou:2025jcv}.}
    \label{fig:gapvsgagamma}
\end{figure}

In Fig.~\ref{fig:gapvsgagamma}, we show the bounds in the $g_{ap}$--$g_{a\gamma}$ plane for the specific value $m_a=3\, \mathrm{ neV}$, representative of the strongest enhancement by turbulence. Our results are leading constraints for $6\times 10^{-15}\, \mathrm{ GeV}^{-1}\lesssim g_{a\gamma}\lesssim 2\times 10^{-12}\, \mathrm{ GeV}^{-1}$ and $3\times 10^{-12}\lesssim g_{ap}\lesssim 1\times 10^{-9}$. The bound flattens out for $g_{ap}\lesssim g_{a\gamma} m_p$, where the flux from Primakoff production prevails over bremsstrahlung~\cite{Fiorillo:2025gnd}.

The bound on $g_{a\gamma}$ shown in Fig.~\ref{fig:gapvsgagamma} arises from a magnetic white dwarf with a particularly low linear polarization signal amplitude~\cite{Benabou:2025jcv}, whereas the bound on $g_{ap}$ from the cooling of SN~1987A \cite{Raffelt:1996wa, Fiorillo:2025gnd}. The product of these individual limits implies $g_{ap}\times g_{a\gamma}\lesssim 2\times 10^{-21}\,\rm GeV^{-1}$ and as such is the main competing bound to those of Fig.~\ref{fig:BremssBounds}. It is comparable to the one from conversion in the assumed but unmeasured $B$ field of the SN~1987A progenitor. A supplemental probe on $g_{a\gamma}\times g_{aN}$ is offered by the deexcitation of sodium through axion emission during the late-stage evolution of massive stars~\cite{Haxton:2025xqz}, but even optimistic future projections are far less sensitive than the bounds of Fig.~\ref{fig:BremssBounds}.

\textbf{\textit{Discussion---}}The nonobservation of $\gamma$ rays in coincidence with SN~1987A neutrinos has led to restrictive limits on axions. Turbulent fields offer a new target for conversions. A previous work found that the average conversion probability for massless axions is boosted by up to a factor 2~\cite{Carenza:2021alz}. While we recover this result, the constraints are actually less restrictive because turbulent conversions in this regime are highly fluctuating and cannot confidently exclude any new parameter space.

On the other hand, we identify a novel resonant target, offered by small-scale turbulent fields, for the conversion of massive axions. In this way, we extend SN~1987A sensitivity to much larger masses (Fig.~\ref{fig:BremssBounds}), surpassing previous bounds by up to 2 orders of magnitude in \hbox{$g_{ap}\times g_{a\gamma}$}. When axions are produced by Primakoff scattering through their photon coupling alone, corresponding constraints on $g_{a\gamma}$ exist, as shown in the~SupM.

The conversion probability $P_{a\gamma}$ in such random fields arises from the resonance with the undulating modes of the turbulent $B$-field. For axions so light that $k_a\ll k_L$, the resonance takes place for all the $B$-field modes, and the average conversion probability, $\langle P_{a\gamma}(k_a)\rangle$, is maximized and independent of $m_a$. If the axion is so heavy that $k_a^{-1}$ is much shorter than the turbulence scales, $\langle P_{a\gamma}(k_a)\rangle$ is suppressed as $m_a^{-4}$. However, we also find an intermediate regime, where only a fraction of the modes are in resonance, a regime characterized by $k_a$ between the cutoffs of the $B$-field power spectrum. In this regime, $\langle P_{a\gamma}(k_a)\rangle$ is suppressed as $m_a^{-2\beta}$, where the power-law index of the turbulent $B$-field is $\beta=5/3$ for a Kolmogorov spectrum. The intermediate regime features small dispersion around $\langle P_{a\gamma}\rangle$ since the fluctuations are averaged out when integrating over the axion spectrum. What matters most, however, is that in this regime axions maintain resonant conversion over the entire propagation length, so the conversion probability is parametrically enhanced by a rather large factor $L k_a$.

A natural future extension of these results would be to apply our formalism to other kinds of turbulent magnetic fields. For example, the turbulent fields of starburst galaxies such as M82 and M87, recently targeted for axion searches~\cite{Ning:2024eky,Candon:2024eah,Ning:2025tit}, can potentially boost the conversion probability, extending existing bounds on, e.g., $g_{a\gamma}$ and $g_{a\gamma}\times g_{ae}$. Yet other possible targets are stellar surfaces, such as on the Sun~\cite{Ruz:2024gkl}, and on red supergiants (RSG)~\cite{Manzari:2024jns}, the typical progenitors of core-collapse SNe. Stars can be covered by convective cells, where turbulent fields exist, so that existing constraints and projected sensitivities might be parametrically affected.

\textbf{\textit{Acknowledgements---}}We thank Tess Jaffe, Foteini Oikonomou, Javier Reynoso-Cordova, Amit Seta, and Michael Unger for helpful discussions or correspondence. AGM and EV acknowledge support from Italian Ministero dell’Universit\`a e della Ricerca through the FIS 2 Project FIS-2023-01577 (DD n. 23314 10-12-2024, CUP C53C24001460001). DFGF, AGM, and EV acknowledge support by Istituto Nazionale di Fisica Nucleare (INFN) through the Theoretical Astroparticle Physics (TAsP) project. EV acknowledges support from the Italian Ministero dell’Universit\`a e della Ricerca Departments of Excellence grant 2023--2027 ``Quantum Frontier''. GGR acknowledges support by the German Research Foundation (DFG) through the Collaborative Research Centre ``Neutrinos and Dark Matter in Astro- and Particle Physics (NDM),'' Grant No. SFB--1258--283604770, and under Germany’s Excellence Strategy through the Cluster of Excellence ORIGINS EXC--2094--390783311. DFGF acknowledges support from the Italian Ministero dell'Universit\`a e della Ricerca through the FIS 3 project FIS-2024-03087 (DD n. 18010 12-11-2025, CUP E53C25002700001). This Letter is based on work from COST Action COSMIC WISPers (CA21106), supported by COST (European Cooperation in Science and Technology).

\bibliography{refs}

@article{Armstrong:1995zc,
    author = "Armstrong, J. W. and Rickett, B. J. and Spangler, S. R.",
    title = "{Electron density power spectrum in the local interstellar medium}",
    doi = "10.1086/175515",
    journal = "Astrophys. J.",
    volume = "443",
    pages = "209--221",
    year = "1995"
}

@article{kolmogorov1991local,
  title="{The local structure of turbulence in incompressible viscous fluid for very large Reynolds numbers}",
  author={Kolmogorov, Andrei Nikolaevich},
  journal = {Akademiia Nauk SSSR Doklady},
         year = 1941,
       volume = {30},
        pages = {299–303},
note="For English translations of the Russian original see
\href{https://doi.org/10.1070/PU1968v010n06ABEH003710}{{\em Sov. Phys. Usp.} {\bf 10} (1968) 734--736}
or
\href{https://doi.org/10.1098/rspa.1991.0075}{{\em Proc. R. Soc. Lond. A} {\bf 434} (1991) 9--13}"
}

@article{Goldreich:1994zz,
    author = "Goldreich, P. and Sridhar, S.",
    title = "{Toward a theory of interstellar turbulence. II. Strong Alfv\'enic turbulence}",
    reportNumber = "CITA-94-26",
    doi = "10.1086/175121",
    journal = "Astrophys. J.",
    volume = "438",
    pages = "763--775",
    year = "1995"
}

@article{Unger:2025xkk,
    author = "Unger, Michael and Farrar, Glennys R.",
    title = "{The Galactic Magnetic Field and UHECR Deflections}",
    eprint = "2502.15876",
    archivePrefix = "arXiv",
    primaryClass = "astro-ph.HE",
    doi = "10.22323/1.484.0003",
    journal = "PoS",
    volume = "UHECR2024",
    pages = "003",
    year = "2025"
}

@article{Korochkin:2024yit,
    author = "Korochkin, Alexander and Semikoz, Dmitri and Tinyakov, Peter",
    title = "{The coherent magnetic field of the Milky Way halo, the Local Bubble, and the Fan region}",
    eprint = "2407.02148",
    archivePrefix = "arXiv",
    primaryClass = "astro-ph.GA",
    doi = "10.1051/0004-6361/202451440",
    journal = "Astron. Astrophys.",
    volume = "693",
    pages = "A284",
    year = "2025"
}

@article{Haverkorn:2008tb,
    author = "Haverkorn, M. and Brown, J. C. and Gaensler, B. M. and McClure-Griffiths, N. M.",
    title = "{The outer scale of turbulence in the magneto-ionized Galactic interstellar medium}",
    eprint = "0802.2740",
    archivePrefix = "arXiv",
    primaryClass = "astro-ph",
    doi = "10.1086/587165",
    journal = "Astrophys. J.",
    volume = "680",
    pages = "362",
    year = "2008"
}

@article{Mertsch:2013pua,
    author = "Mertsch, Philipp and Sarkar, Subir",
    title = "{Loops and spurs: The angular power spectrum of the Galactic synchrotron background}",
    eprint = "1304.1078",
    archivePrefix = "arXiv",
    primaryClass = "astro-ph.GA",
    doi = "10.1088/1475-7516/2013/06/041",
    journal = "JCAP",
    volume = "06",
    pages = "041",
    year = "2013"
}

@article{Beck:2014pma,
    author = "Beck, Marcus C. and Beck, Alexander M. and Beck, Rainer and Dolag, Klaus and Strong, Andrew W. and Nielaba, Peter",
    title = "{New constraints on modelling the random magnetic field of the MW}",
    eprint = "1409.5120",
    archivePrefix = "arXiv",
    primaryClass = "astro-ph.GA",
    doi = "10.1088/1475-7516/2016/05/056",
    journal = "JCAP",
    volume = "05",
    pages = "056",
    year = "2016"
}

@article{Carenza:2022zmq,
    author = "Carenza, Pierluca and Sharma, Ramkishor and Marsh, M. C. David and Brandenburg, Axel and Ravensburg, Eike",
    title = "{Magnetohydrodynamics predicts heavy-tailed distributions of axion-photon conversion}",
    eprint = "2208.04333",
    archivePrefix = "arXiv",
    primaryClass = "hep-ph",
    reportNumber = "NORDITA 2022-060",
    doi = "10.1103/PhysRevD.108.103029",
    journal = "Phys. Rev. D",
    volume = "108",
    number = "10",
    pages = "103029",
    year = "2023"
}

@ARTICLE{2019Natur.567..200P,
       author = {{Pietrzy{\'n}ski}, G. and others},
        title = "{A distance to the Large Magellanic Cloud that is precise to one per cent}",
      journal = {Nature},
         year = 2019,
        month = mar,
       volume = {567},
       number = {7747},
        pages = {200-203},
          doi = {10.1038/s41586-019-0999-4},
archivePrefix = {arXiv},
       eprint = {1903.08096},
 primaryClass = {astro-ph.GA},
       adsurl = {https://ui.adsabs.harvard.edu/abs/2019Natur.567..200P}
}

@misc{Panagia:1998,
       author = {Panagia, N.},
       title = "{Distance to SN~1987A and the LMC}",
       note = "{in: New Views of the Magellanic Clouds,
       Proc.\ IAU Symposium No.\ 190, Victoria, Canada, 12--17 July 1998, eds.\ Y.-H.~Chu et al., Astron.\ Soc.\ of the Pacific (1999), 549--553. ADS Link:
       \href{https://articles.adsabs.harvard.edu/pdf/1999IAUS..190..549P}{https://articles.adsabs.harvard.edu /pdf/1999IAUS..190..549P}}"
}

@ARTICLE{Panagia:2003,
       author = {{Panagia}, Nino},
        title = "{A Geometric Determination of the Distance to SN 1987A and the LMC}",
         year = 2003,
       eprint = {astro-ph/0309416},
      note = "{in: IAU Colloq.\ 192: Cosmic Explosions, On the 10th Anniversary of SN1993J, J.-M.\ Marcaide and K.~W.\ Weiler (eds.), \href{https://doi.org/10.1007/3-540-26633-X_78}{{\em Springer Proc.\ Phys.} {\bf 99} (2005) 585}}"
}

@book{padmanabhan2000theoretical,
  title={Theoretical Astrophysics. Vol. 1: Astrophysical Processes},
  author={Padmanabhan, P},
  year={2000},
  publisher = {Cambridge University Press}
}

@article{Ning:2025tit,
    author = "Ning, Orion and Safdi, Benjamin R.",
    title = "{Probing the axion-electron coupling with NuSTAR observations of galaxies}",
    eprint = "2503.09682",
    archivePrefix = "arXiv",
    primaryClass = "hep-ph",
    doi = "10.1103/28vj-3rws",
    journal = "Phys. Rev. D",
    volume = "113",
    number = "1",
    pages = "015021",
    year = "2026"
}

@article{Marsh:2021ajy,
    author = "Marsh, M. C. David and Matthews, James H. and Reynolds, Christopher and Carenza, Pierluca",
    title = "{Fourier formalism for relativistic axion-photon conversion with astrophysical applications}",
    eprint = "2107.08040",
    archivePrefix = "arXiv",
    primaryClass = "hep-ph",
    doi = "10.1103/PhysRevD.105.016013",
    journal = "Phys. Rev. D",
    volume = "105",
    number = "1",
    pages = "016013",
    year = "2022"
}

@article{Arza:2026rsl,
    author = "Arza, A. and others",
    title = "{The COSMIC WISPers White Paper: The physics case for Weakly Interacting Slim Particles}",
    eprint = "2603.03433",
    archivePrefix = "arXiv",
    primaryClass = "hep-ph",
    reportNumber = "BARI-TH/784-26, CERN-TH-2026-016, IPPP/26/13, IFT-UAM/CSIC-26-13, KCL-PH-TH/2026-04, KEK-Cosmo-0411, KEK-TH-2804, LAPTH-008/26, MPP-2026-21, RESCEU-5/26, SLAC-PUB-260219, ST/T006994/1, ST/Y004531/1",
    month = "3",
    year = "2026"
}

@article{Raffelt:2025wty,
    author = "Raffelt, Georg G. and Janka, Hans-Thomas and Fiorillo, Damiano F. G.",
    title = "{Neutrinos from core-collapse supernovae}",
    eprint = "2509.16306",
    archivePrefix = "arXiv",
    year = "2025",
    note="To be published in the Encyclopedia of Particle Physics"
}

@article{vanBibber:1988ge,
    author = "van Bibber, K. and McIntyre, P. M. and Morris, D. E. and Raffelt, G. G.",
    title = "{A Practical Laboratory Detector for Solar Axions}",
    reportNumber = "LBL-25908",
    doi = "10.1103/PhysRevD.39.2089",
    journal = "Phys. Rev. D",
    volume = "39",
    pages = "2089",
    year = "1989"
}

@article{VanBibber:1987rq,
    author = "Van Bibber, K. and Dagdeviren, N. R. and Koonin, S. E. and Kerman, A. and Nelson, H. N.",
    title = "{Proposed experiment to produce and detect light pseudoscalars}",
    reportNumber = "SLAC-PUB-4322",
    doi = "10.1103/PhysRevLett.59.759",
    journal = "Phys. Rev. Lett.",
    volume = "59",
    pages = "759--762",
    year = "1987"
}

@article{Irastorza:2018dyq,
    author = "Irastorza, Igor G. and Redondo, Javier",
    title = "{New experimental approaches in the search for axion-like particles}",
    eprint = "1801.08127",
    archivePrefix = "arXiv",
    primaryClass = "hep-ph",
    doi = "10.1016/j.ppnp.2018.05.003",
    journal = "Prog. Part. Nucl. Phys.",
    volume = "102",
    pages = "89--159",
    year = "2018"
}

@article{Baryakhtar:2025jwh,
    author = "Baryakhtar, Masha and Rosenberg, Leslie and Rybka, Gray",
    title = "{Searching for the QCD Dark Matter Axion}",
    eprint = "2504.10607",
    archivePrefix = "arXiv",
    primaryClass = "hep-ex",
    month = "4",
    year = "2025"
}

@article{Kamiokande-II:1987idp,
    author = "Hirata, K. and others",
    editor = "Wali, K. C.",
    collaboration = "Kamiokande-II",
    title = "{Observation of a Neutrino Burst from the Supernova SN 1987a}",
    reportNumber = "UT-ICEPP-87-01, UPR-142E",
    doi = "10.1103/PhysRevLett.58.1490",
    journal = "Phys. Rev. Lett.",
    volume = "58",
    pages = "1490--1493",
    year = "1987"
}

@article{Hirata:1988ad,
    author = "Hirata, K. S. and others",
    title = "{Observation in the Kamiokande-II Detector of the Neutrino Burst from Supernova SN 1987a}",
    doi = "10.1103/PhysRevD.38.448",
    journal = "Phys. Rev. D",
    volume = "38",
    pages = "448--458",
    year = "1988"
}

@article{Bionta:1987qt,
    author = "Bionta, R. M. and others",
    title = "{Observation of a Neutrino Burst in Coincidence with Supernova SN 1987a in the Large Magellanic Cloud}",
    reportNumber = "UCI-NEUTRINO-87-10",
    doi = "10.1103/PhysRevLett.58.1494",
    journal = "Phys. Rev. Lett.",
    volume = "58",
    pages = "1494",
    year = "1987"
}

@article{IMB:1988suc,
    author = "Bratton, C. B. and others",
    collaboration = "IMB",
    title = "{Angular Distribution of Events From Sn1987a}",
    reportNumber = "UM-PDK-88-1",
    doi = "10.1103/PhysRevD.37.3361",
    journal = "Phys. Rev. D",
    volume = "37",
    pages = "3361",
    year = "1988"
}

@article{Alekseev:1987ej,
    author = "Alekseev, E. N. and Alekseeva, L. N. and Volchenko, V. I. and Krivosheina, I. V.",
    editor = "Tran Thanh Van, J.",
    title = "{Possible Detection of a Neutrino Signal on 23 February 1987 at the Baksan Underground Scintillation Telescope of the Institute of Nuclear Research}",
    journal = "JETP Lett.",
    volume = "45",
    pages = "589--592",
    year = "1987"
}

@article{Alekseev:1988gp,
    author = "Alekseev, E. N. and Alekseeva, L. N. and Krivosheina, I. V. and Volchenko, V. I.",
    title = "{Detection of the Neutrino Signal From {SN1987A} in the {LMC} Using the Inr Baksan Underground Scintillation Telescope}",
    doi = "10.1016/0370-2693(88)91651-6",
    journal = "Phys. Lett. B",
    volume = "205",
    pages = "209--214",
    year = "1988"
}

@article{Koshiba:1992yb,
    author = "Koshiba, M.",
    title = "{Observational neutrino astrophysics}",
    doi = "10.1016/0370-1573(92)90083-C",
    journal = "Phys. Rept.",
    volume = "220",
    pages = "229--381",
    year = "1992"
}

@article{Mayle:1987as,
    author = "Mayle, Ron and Wilson, James R. and Ellis, John R. and Olive, Keith A. and Schramm, David N. and Steigman, Gary",
    title = "{Constraints on Axions from SN 1987A}",
    reportNumber = "FERMILAB-PUB-87-225-A, EFI-87-104-CHICAGO, UMN-TH-637-87, CERN-TH-4887-87",
    doi = "10.1016/0370-2693(88)91595-X",
    journal = "Phys. Lett. B",
    volume = "203",
    pages = "188--196",
    year = "1988"
}

@article{Turner:1987by,
    author = "Turner, Michael S.",
    title = "{Axions from SN 1987A}",
    reportNumber = "FERMILAB-PUB-87-202-A",
    doi = "10.1103/PhysRevLett.60.1797",
    journal = "Phys. Rev. Lett.",
    volume = "60",
    pages = "1797",
    year = "1988"
}

@article{Burrows:1988ah,
    author = "Burrows, Adam and Turner, Michael S. and Brinkmann, R. P.",
    title = "{Axions and SN 1987A}",
    reportNumber = "FERMILAB-PUB-88-105-A",
    doi = "10.1103/PhysRevD.39.1020",
    journal = "Phys. Rev. D",
    volume = "39",
    pages = "1020",
    year = "1989"
}

@article{Mayle:1989yx,
    author = "Mayle, Ron and Wilson, James R. and Ellis, John R. and Olive, Keith A. and Schramm, David N. and Steigman, Gary",
    title = "{Updated Constraints on Axions from SN 1987A}",
    reportNumber = "PRINT-89-0012 (OHIO-STATE), FERMILAB-PUB-88-209-A",
    doi = "10.1016/0370-2693(89)91104-0",
    journal = "Phys. Lett. B",
    volume = "219",
    pages = "515",
    year = "1989"
}

@article{Burrows:1990pk,
    author = "Burrows, Adam and Ressell, M. Ted and Turner, Michael S.",
    title = "{Axions and SN~1987A: Axion Trapping}",
    reportNumber = "FERMILAB-PUB-90-081-A, FERMILAB-PUB-90-081-A-REV",
    doi = "10.1103/PhysRevD.42.3297",
    journal = "Phys. Rev. D",
    volume = "42",
    pages = "3297--3309",
    year = "1990"
}

@article{forrest1980gamma,
  title={The gamma ray spectrometer for the Solar Maximum Mission},
  author={Forrest, DJ and Chupp, EL and Ryan, JM and Cherry, ML and Gleske, IU and Reppin, C and Pinkau, K and Rieger, E and Kanbach, G and Kinzer, RL and others},
  journal={Solar Physics},
  volume={65},
  number={1},
  pages={15--23},
  year={1980},
  publisher={Springer}
}

@article{Chupp:1989kx,
    author = "Chupp, E. L. and Vestrand, W. T. and Reppin, C.",
    title = "{Experimental Limits on the Radiative Decay of SN~1987A Neutrinos}",
    doi = "10.1103/PhysRevLett.62.505",
    journal = "Phys. Rev. Lett.",
    volume = "62",
    pages = "505--508",
    year = "1989"
}

@article{Kolb:1988pe,
    author = "Kolb, Edward W. and Turner, Michael S.",
    title = "{Limits to the Radiative Decays of Neutrinos and Axions from Gamma-Ray Observations of SN 1987A}",
    reportNumber = "FERMILAB-PUB-87-223-A, FERMILAB-PUB-87-223-A-REV",
    doi = "10.1103/PhysRevLett.62.509",
    journal = "Phys. Rev. Lett.",
    volume = "62",
    pages = "509",
    year = "1989"
}

@article{Brockway:1996yr,
    author = "Brockway, Jack W. and Carlson, Eric D. and Raffelt, Georg G.",
    title = "{SN~1987A gamma-ray limits on the conversion of pseudoscalars}",
    eprint = "astro-ph/9605197",
    archivePrefix = "arXiv",
    reportNumber = "MPI-PHT-96-42",
    doi = "10.1016/0370-2693(96)00778-2",
    journal = "Phys. Lett. B",
    volume = "383",
    pages = "439--443",
    year = "1996"
}

@article{Grifols:1996id,
    author = "Grifols, J. A. and Masso, E. and Toldra, R.",
    title = "{Gamma-rays from SN~1987A due to pseudoscalar conversion}",
    eprint = "astro-ph/9606028",
    archivePrefix = "arXiv",
    reportNumber = "UAB-FT-391",
    doi = "10.1103/PhysRevLett.77.2372",
    journal = "Phys. Rev. Lett.",
    volume = "77",
    pages = "2372--2375",
    year = "1996"
}

@article{Payez:2014xsa,
    author = "Payez, Alexandre and Evoli, Carmelo and Fischer, Tobias and Giannotti, Maurizio and Mirizzi, Alessandro and Ringwald, Andreas",
    title = "{Revisiting the SN1987A gamma-ray limit on ultralight axion-like particles}",
    eprint = "1410.3747",
    archivePrefix = "arXiv",
    primaryClass = "astro-ph.HE",
    reportNumber = "DESY-14-164",
    doi = "10.1088/1475-7516/2015/02/006",
    journal = "JCAP",
    volume = "02",
    pages = "006",
    year = "2015"
}

@article{Hoof:2022xbe,
    author = "Hoof, Sebastian and Schulz, Lena",
    title = "{Updated constraints on axion-like particles from temporal information in supernova SN1987A gamma-ray data}",
    eprint = "2212.09764",
    archivePrefix = "arXiv",
    primaryClass = "hep-ph",
    reportNumber = "TTP22-072",
    doi = "10.1088/1475-7516/2023/03/054",
    journal = "JCAP",
    volume = "03",
    pages = "054",
    year = "2023"
}

@article{Candon:2024eah,
    author = "Cand{\'o}n, Francisco R. and Fiorillo, Damiano F. G. and Lucente, Giuseppe and Vitagliano, Edoardo and Vogel, Julia K.",
    title = "{NuSTAR Bounds on Radiatively Decaying Particles from M82}",
    eprint = "2412.03660",
    archivePrefix = "arXiv",
    primaryClass = "hep-ph",
    doi = "10.1103/PhysRevLett.134.171004",
    journal = "Phys. Rev. Lett.",
    volume = "134",
    number = "17",
    pages = "171004",
    year = "2025"
}

@article{Svrcek:2006yi,
    author = "Svrcek, Peter and Witten, Edward",
    title = "{Axions In String Theory}",
    eprint = "hep-th/0605206",
    archivePrefix = "arXiv",
    reportNumber = "SLAC-PUB-11894",
    doi = "10.1088/1126-6708/2006/06/051",
    journal = "JHEP",
    volume = "06",
    pages = "051",
    year = "2006"
}

@article{Arvanitaki:2009fg,
    author = "Arvanitaki, Asimina and Dimopoulos, Savas and Dubovsky, Sergei and Kaloper, Nemanja and March-Russell, John",
    title = "{String Axiverse}",
    eprint = "0905.4720",
    archivePrefix = "arXiv",
    primaryClass = "hep-th",
    doi = "10.1103/PhysRevD.81.123530",
    journal = "Phys. Rev. D",
    volume = "81",
    pages = "123530",
    year = "2010"
}

@article{Cicoli:2012sz,
    author = "Cicoli, Michele and Goodsell, Mark and Ringwald, Andreas",
    title = "{The type IIB string axiverse and its low-energy phenomenology}",
    eprint = "1206.0819",
    archivePrefix = "arXiv",
    primaryClass = "hep-th",
    reportNumber = "DESY-12-058, CERN-PH-TH-2012-153",
    doi = "10.1007/JHEP10(2012)146",
    journal = "JHEP",
    volume = "10",
    pages = "146",
    year = "2012"
}

@article{Demirtas:2018akl,
    author = "Demirtas, Mehmet and Long, Cody and McAllister, Liam and Stillman, Mike",
    title = "{The Kreuzer-Skarke Axiverse}",
    eprint = "1808.01282",
    archivePrefix = "arXiv",
    primaryClass = "hep-th",
    doi = "10.1007/JHEP04(2020)138",
    journal = "JHEP",
    volume = "04",
    pages = "138",
    year = "2020"
}

@article{Halverson:2019cmy,
    author = "Halverson, James and Long, Cody and Nelson, Brent and Salinas, Gustavo",
    title = "{Towards string theory expectations for photon couplings to axionlike particles}",
    eprint = "1909.05257",
    archivePrefix = "arXiv",
    primaryClass = "hep-th",
    doi = "10.1103/PhysRevD.100.106010",
    journal = "Phys. Rev. D",
    volume = "100",
    number = "10",
    pages = "106010",
    year = "2019"
}

@article{Demirtas:2021gsq,
    author = "Demirtas, Mehmet and Gendler, Naomi and Long, Cody and McAllister, Liam and Moritz, Jakob",
    title = "{PQ axiverse}",
    eprint = "2112.04503",
    archivePrefix = "arXiv",
    primaryClass = "hep-th",
    doi = "10.1007/JHEP06(2023)092",
    journal = "JHEP",
    volume = "06",
    pages = "092",
    year = "2023"
}

@article{Gendler:2023kjt,
    author = "Gendler, Naomi and Marsh, David J. E. and McAllister, Liam and Moritz, Jakob",
    title = "{Glimmers from the axiverse}",
    eprint = "2309.13145",
    archivePrefix = "arXiv",
    primaryClass = "hep-th",
    reportNumber = "KCL-PH-TH/2023-49",
    doi = "10.1088/1475-7516/2024/09/071",
    journal = "JCAP",
    volume = "09",
    pages = "071",
    year = "2024"
}

@article{Candon:2025sdm,
    author = "Cand{\'o}n, Francisco R. and Fiorillo, Damiano F. G. and Gil Muyor, {\'A}ngel and Janka, Hans-Thomas and Raffelt, Georg G. and Vitagliano, Edoardo",
    title = "{Stripped-Envelope Supernovae for QCD Axion Detection}",
    eprint = "2511.13815",
    archivePrefix = "arXiv",
    primaryClass = "hep-ph",
    doi = "10.1103/bqpl-zyr8",
    journal = "Phys. Rev. Lett.",
    volume = "136",
    number = "17",
    pages = "171001",
    year = "2026"
}

@article{Kim:1979if,
    author = "Kim, Jihn E.",
    title = "{Weak Interaction Singlet and Strong CP Invariance}",
    reportNumber = "UPR-0120T",
    doi = "10.1103/PhysRevLett.43.103",
    journal = "Phys. Rev. Lett.",
    volume = "43",
    pages = "103",
    year = "1979"
}

@article{Shifman:1979if,
    author = "Shifman, Mikhail A. and Vainshtein, A. I. and Zakharov, Valentin I.",
    title = "{Can Confinement Ensure Natural CP Invariance of Strong Interactions?}",
    reportNumber = "ITEP-64-1979",
    doi = "10.1016/0550-3213(80)90209-6",
    journal = "Nucl. Phys. B",
    volume = "166",
    pages = "493--506",
    year = "1980"
}

@article{Dine:1981rt,
    author = "Dine, Michael and Fischler, Willy and Srednicki, Mark",
    title = "{A Simple Solution to the Strong CP Problem with a Harmless Axion}",
    reportNumber = "Print-81-0320 (IAS,PRINCETON)",
    doi = "10.1016/0370-2693(81)90590-6",
    journal = "Phys. Lett. B",
    volume = "104",
    pages = "199--202",
    year = "1981"
}

@article{Zhitnitsky:1980tq,
    author = "Zhitnitsky, A. R.",
    title = "{On Possible Suppression of the Axion Hadron Interactions}",
    journal = "Sov. J. Nucl. Phys.",
    volume = "31",
    pages = "260",
    year = "1980",
    note = "Translated from {\em Yad. Fiz.} {\bf 31} (1980) 497"
}

@article{Sikivie:2020zpn,
    author = "Sikivie, Pierre",
    title = "{Invisible Axion Search Methods}",
    eprint = "2003.02206",
    archivePrefix = "arXiv",
    primaryClass = "hep-ph",
    doi = "10.1103/RevModPhys.93.015004",
    journal = "Rev. Mod. Phys.",
    volume = "93",
    number = "1",
    pages = "015004",
    year = "2021"
}

@article{Unger:2023lob,
    author = "Unger, Michael and Farrar, Glennys R.",
    title = "{The Coherent Magnetic Field of the Milky Way}",
    eprint = "2311.12120",
    archivePrefix = "arXiv",
    primaryClass = "astro-ph.GA",
    doi = "10.3847/1538-4357/ad4a54",
    journal = "Astrophys. J.",
    volume = "970",
    number = "1",
    pages = "95",
    year = "2024"
}

@book{Raffelt:1996wa,
    author = "Raffelt, G. G.",
    title = "{Stars as Laboratories for Fundamental Physics}",
    isbn = "978-0-226-70272-8",
    year = "1996",
    publisher="{University of Chicago Press}"
}

@article{Fiorillo:2023frv,
    author = "Fiorillo, Damiano F. G. and Heinlein, Malte and Janka, Hans-Thomas and Raffelt, Georg and Vitagliano, Edoardo and Bollig, Robert",
    title = "{Supernova simulations confront SN 1987A neutrinos}",
    eprint = "2308.01403",
    archivePrefix = "arXiv",
    primaryClass = "astro-ph.HE",
    doi = "10.1103/PhysRevD.108.083040",
    journal = "Phys. Rev. D",
    volume = "108",
    number = "8",
    pages = "083040",
    year = "2023"
}

@article{Peccei:1977hh,
    author = "Peccei, R. D. and Quinn, Helen R.",
    title = "{CP Conservation in the Presence of Instantons}",
    reportNumber = "ITP-568-STANFORD",
    doi = "10.1103/PhysRevLett.38.1440",
    journal = "Phys. Rev. Lett.",
    volume = "38",
    pages = "1440--1443",
    year = "1977"
}

@article{Weinberg:1977ma,
    author = "Weinberg, Steven",
    title = "{A New Light Boson?}",
    reportNumber = "HUTP-77/A074",
    doi = "10.1103/PhysRevLett.40.223",
    journal = "Phys. Rev. Lett.",
    volume = "40",
    pages = "223--226",
    year = "1978"
}

@article{Wilczek:1977pj,
    author = "Wilczek, Frank",
    title = "{Problem of Strong  $P$  and  $T$  Invariance in the Presence of Instantons}",
    reportNumber = "Print-77-0939 (COLUMBIA)",
    doi = "10.1103/PhysRevLett.40.279",
    journal = "Phys. Rev. Lett.",
    volume = "40",
    pages = "279--282",
    year = "1978"
}

@article{Preskill:1982cy,
    author = "Preskill, John and Wise, Mark B. and Wilczek, Frank",
    editor = "Srednicki, M. A.",
    title = "{Cosmology of the Invisible Axion}",
    reportNumber = "HUTP-82-A048, NSF-ITP-82-103",
    doi = "10.1016/0370-2693(83)90637-8",
    journal = "Phys. Lett. B",
    volume = "120",
    pages = "127--132",
    year = "1983"
}

@article{Abbott:1982af,
    author = "Abbott, L. F. and Sikivie, P.",
    editor = "Srednicki, M. A.",
    title = "{A Cosmological Bound on the Invisible Axion}",
    reportNumber = "PRINT-82-0695 (BRANDEIS)",
    doi = "10.1016/0370-2693(83)90638-X",
    journal = "Phys. Lett. B",
    volume = "120",
    pages = "133--136",
    year = "1983"
}

@article{Peccei:1977ur,
    author = "Peccei, R. D. and Quinn, Helen R.",
    title = "{Constraints Imposed by CP Conservation in the Presence of Instantons}",
    reportNumber = "ITP-572-STANFORD",
    doi = "10.1103/PhysRevD.16.1791",
    journal = "Phys. Rev. D",
    volume = "16",
    pages = "1791--1797",
    year = "1977"
}

@article{Dine:1982ah,
    author = "Dine, Michael and Fischler, Willy",
    editor = "Srednicki, M. A.",
    title = "{The Not So Harmless Axion}",
    reportNumber = "UPR-0201T",
    doi = "10.1016/0370-2693(83)90639-1",
    journal = "Phys. Lett. B",
    volume = "120",
    pages = "137--141",
    year = "1983"
}

@article{OHare:2024nmr,
    author = "O'Hare, Ciaran A. J.",
    title = "{Cosmology of axion dark matter}",
    eprint = "2403.17697",
    archivePrefix = "arXiv",
    primaryClass = "hep-ph",
    doi = "10.22323/1.454.0040",
    journal = "PoS",
    volume = "COSMICWISPers",
    pages = "040",
    year = "2024"
}

@article{DiLuzio:2020wdo,
    author = "Di Luzio, Luca and Giannotti, Maurizio and Nardi, Enrico and Visinelli, Luca",
    title = "{The landscape of QCD axion models}",
    eprint = "2003.01100",
    archivePrefix = "arXiv",
    primaryClass = "hep-ph",
    reportNumber = "DESY 20-036, DESY-20-036",
    doi = "10.1016/j.physrep.2020.06.002",
    journal = "Phys. Rept.",
    volume = "870",
    pages = "1--117",
    year = "2020"
}

@article{Oberauer:1993yr,
    author = "Oberauer, L. and Hagner, C. and Raffelt, G. and Rieger, E.",
    title = "{Supernova bounds on neutrino radiative decays}",
    doi = "10.1016/0927-6505(93)90004-W",
    journal = "Astropart. Phys.",
    volume = "1",
    pages = "377--386",
    year = "1993"
}

@article{Raffelt:2006cw,
    author = "Raffelt, Georg G.",
    editor = "Kuster, Markus and Raffelt, Georg and Beltran, Berta",
    title = "{Astrophysical axion bounds}",
    eprint = "hep-ph/0611350",
    archivePrefix = "arXiv",
    reportNumber = "MPP-2006-172",
    doi = "10.1007/978-3-540-73518-2_3",
    journal = "Lect. Notes Phys.",
    volume = "741",
    pages = "51--71",
    year = "2008"
}

@article{Lella:2024hfk,
    author = "Lella, Alessandro and Calore, Francesca and Carenza, Pierluca and Eckner, Christopher and Giannotti, Maurizio and Lucente, Giuseppe and Mirizzi, Alessandro",
    title = "{Probing protoneutron stars with gamma-ray axionscopes}",
    eprint = "2405.02395",
    archivePrefix = "arXiv",
    primaryClass = "hep-ph",
    reportNumber = "LAPTH-024/24, BARI-TH/769-24",
    doi = "10.1088/1475-7516/2024/11/009",
    journal = "JCAP",
    volume = "11",
    pages = "009",
    year = "2024"
}

@article{Graham:2015ouw,
    author = "Graham, Peter W. and Irastorza, Igor G. and Lamoreaux, Steven K. and Lindner, Axel and van Bibber, Karl A.",
    title = "{Experimental Searches for the Axion and Axion-Like Particles}",
    eprint = "1602.00039",
    archivePrefix = "arXiv",
    primaryClass = "hep-ex",
    doi = "10.1146/annurev-nucl-102014-022120",
    journal = "Ann. Rev. Nucl. Part. Sci.",
    volume = "65",
    pages = "485--514",
    year = "2015"
}

@article{CAST:2024eil,
    author = {Altenm\"uller, K. and others},
    collaboration = "CAST",
    title = "{New Upper Limit on the Axion-Photon Coupling with an Extended CAST Run with a Xe-Based Micromegas Detector}",
    eprint = "2406.16840",
    archivePrefix = "arXiv",
    primaryClass = "hep-ex",
    doi = "10.1103/PhysRevLett.133.221005",
    journal = "Phys. Rev. Lett.",
    volume = "133",
    number = "22",
    pages = "221005",
    year = "2024"
}

@article{Ruz:2024gkl,
    author = "Ruz, J. and others",
    title = "{NuSTAR as an Axion Helioscope}",
    eprint = "2407.03828",
    archivePrefix = "arXiv",
    primaryClass = "astro-ph.CO",
    doi = "10.1103/18sn-hxtb",
    journal = "Phys. Rev. Lett.",
    volume = "135",
    number = "14",
    pages = "141001",
    year = "2025"
}

@article{Noordhuis:2022ljw,
    author = "Noordhuis, Dion and Prabhu, Anirudh and Witte, Samuel J. and Chen, Alexander Y. and Cruz, F{\'a}bio and Weniger, Christoph",
    title = "{Novel Constraints on Axions Produced in Pulsar Polar-Cap Cascades}",
    eprint = "2209.09917",
    archivePrefix = "arXiv",
    primaryClass = "hep-ph",
    doi = "10.1103/PhysRevLett.131.111004",
    journal = "Phys. Rev. Lett.",
    volume = "131",
    number = "11",
    pages = "111004",
    year = "2023"
}

@article{Manzari:2024jns,
    author = "Manzari, Claudio Andrea and Park, Yujin and Safdi, Benjamin R. and Savoray, Inbar",
    title = "{Supernova Axions Convert to Gamma Rays in Magnetic Fields of Progenitor Stars}",
    eprint = "2405.19393",
    archivePrefix = "arXiv",
    primaryClass = "hep-ph",
    doi = "10.1103/PhysRevLett.133.211002",
    journal = "Phys. Rev. Lett.",
    volume = "133",
    number = "21",
    pages = "211002",
    year = "2024"
}

@article{Carenza:2019pxu,
    author = "Carenza, Pierluca and Fischer, Tobias and Giannotti, Maurizio and Guo, Gang and Mart\'\i{}nez-Pinedo, Gabriel and Mirizzi, Alessandro",
    title = "{Improved axion emissivity from a supernova via nucleon-nucleon bremsstrahlung}",
    eprint = "1906.11844",
    archivePrefix = "arXiv",
    primaryClass = "hep-ph",
    doi = "10.1088/1475-7516/2019/10/016",
    journal = "JCAP",
    volume = "10",
    number = "10",
    pages = "016",
    year = "2019",
    note = "Erratum \href{https://doi.org/10.1088/1475-7516/2020/05/E01}{{\em JCAP} {\bf 05} (2020) E01}"
}

@article{Sikivie:1983ip,
    author = "Sikivie, P.",
    title = "{Experimental Tests of the Invisible Axion}",
    reportNumber = "PRINT-83-0597 (FLORIDA), UF-TP-83-13",
    doi = "10.1103/PhysRevLett.51.1415",
    journal = "Phys. Rev. Lett.",
    volume = "51",
    pages = "1415--1417",
    year = "1983",
    note = "Erratum \href{https://doi.org/10.1103/PhysRevLett.52.695.2}{{\em Phys. Rev. Lett.} {\bf 52} (1984) 695}"
}

@article{Raffelt:1987im,
    author = "Raffelt, Georg and Stodolsky, Leo",
    title = "{Mixing of the Photon with Low Mass Particles}",
    reportNumber = "MPI-PAE/PTh-54/87",
    doi = "10.1103/PhysRevD.37.1237",
    journal = "Phys. Rev. D",
    volume = "37",
    pages = "1237",
    year = "1988"
}

@misc{OHare,
author="O'Hare, Ciaran",
note= "Github Page \href{https://github.com/cajohare}{https://github.com/cajohare}"
}

@article{Haxton:2025xqz,
    author = "Haxton, W. C. and Liu, Xing and McCutcheon, Annie and Ray, Anupam",
    title = "{A Continuous Galactic Line Source of Axions: The Remarkable Case of $^{23}Na$}",
    eprint = "2505.03038",
    archivePrefix = "arXiv",
    primaryClass = "astro-ph.HE",
    doi = "10.1103/n3v9-2wml",
    journal = "Phys. Rev. Lett.",
    volume = "135",
    number = "22",
    pages = "222701",
    year = "2025"
}

@article{Ning:2024eky,
    author = "Ning, Orion and Safdi, Benjamin R.",
    title = "{Leading Axion-Photon Sensitivity with NuSTAR Observations of M82 and M87}",
    eprint = "2404.14476",
    archivePrefix = "arXiv",
    primaryClass = "hep-ph",
    doi = "10.1103/PhysRevLett.134.171003",
    journal = "Phys. Rev. Lett.",
    volume = "134",
    number = "17",
    pages = "171003",
    year = "2025"
}

@article{Benabou:2025jcv,
    author = "Benabou, Joshua N. and Dessert, Christopher and Patra, Kishore C. and Brink, Thomas G. and Zheng, WeiKang and Filippenko, Alexei V. and Safdi, Benjamin R.",
    title = "{Search for Axions in Magnetic White Dwarf Polarization at Lick and Keck Observatories}",
    eprint = "2504.12377",
    archivePrefix = "arXiv",
    primaryClass = "hep-ph",
    month = "4",
    year = "2025"
}

@article{Fiorillo:2025gnd,
    author = "Fiorillo, Damiano F. G. and Gil Muyor, {\'A}ngel and Janka, Hans-Thomas and Raffelt, Georg G. and Vitagliano, Edoardo",
    title = "{Axion-photon conversion in transient compact stars: Systematics, constraints, and opportunities}",
    eprint = "2509.13322",
    archivePrefix = "arXiv",
    primaryClass = "hep-ph",
    doi = "10.1088/1475-7516/2026/03/053",
    journal = "JCAP",
    volume = "03",
    pages = "053",
    year = "2026"
}

@article{Berlin:2024pzi,
    author = "Berlin, Asher and Kahn, Yonatan",
    title = "{New Technologies for Axion and Dark Photon Searches}",
    eprint = "2412.08704",
    archivePrefix = "arXiv",
    primaryClass = "hep-ph",
    reportNumber = "FERMILAB-PUB-24-0933-T",
    doi = "10.1146/annurev-nucl-121423-101015",
    journal = "Ann. Rev. Nucl. Part. Sci.",
    volume = "75",
    number = "1",
    pages = "83--108",
    year = "2025"
}

@article{Mao:2012hd,
    author = "Mao, S. A. and McClure-Griffiths, N. M. and Gaensler, B. M. and Haverkorn, M. and Beck, R. and McConnell, D. and Wolleben, M. and Stanimirovi\'c, S. and Dickey, J. M. and Staveley-Smith, L.",
    title = "{Magnetic Field Structure of the Large Magellanic Cloud from Faraday Rotation Measures of Diffuse Polarized Emission}",
    eprint = "1209.1115",
    archivePrefix = "arXiv",
    primaryClass = "astro-ph.GA",
    doi = "10.1088/0004-637X/759/1/25",
    journal = "Astrophys. J.",
    volume = "759",
    pages = "25",
    year = "2012"
}

@article{Gaensler:2005qj,
    author = "Gaensler, Bryan M. and Haverkorn, M. and Staveley-Smith, L. and Dickey, J. M. and McClure-Griffiths, N. M. and Dickel, J. R. and Wolleben, M.",
    title = "{The Magnetic field of the Large Magellanic Cloud revealed through Faraday rotation}",
    eprint = "astro-ph/0503226",
    archivePrefix = "arXiv",
    doi = "10.1126/science.1108832",
    journal = "Science",
    volume = "307",
    pages = "1610--1612",
    year = "2005"
}

@article{Seta:2022uoy,
    author = "Seta, Amit and Federrath, Christoph and Livingston, Jack D. and McClure-Griffiths, N. M.",
    title = "{Rotation measure structure functions with higher-order stencils as a probe of small-scale magnetic fluctuations and its application to the Small and Large Magellanic Clouds}",
    eprint = "2206.13798",
    archivePrefix = "arXiv",
    primaryClass = "astro-ph.GA",
    doi = "10.1093/mnras/stac2972",
    journal = "Mon. Not. Roy. Astron. Soc.",
    volume = "518",
    number = "1",
    pages = "919--944",
    year = "2022"
}

@article{Reynoso-Cordova:2025meg,
    author = "Reynoso-Cordova, Javier and Gaggero, Daniele and Regis, Marco and Taoso, Marco",
    title = "{Diffusion coefficient in the Large Magellanic Cloud}",
    eprint = "2512.14906",
    archivePrefix = "arXiv",
    primaryClass = "astro-ph.GA",
    doi = "10.1103/rdyk-njx2",
    journal = "Phys. Rev. D",
    volume = "113",
    number = "12",
    pages = "123028",
    year = "2026"
}

@article{Subramanian:2008pi,
    author = "Subramanian, Smitha and Subramaniam, Annapurni",
    title = "{Depth Estimation of the Large and Small Magellanic Clouds}",
    eprint = "0809.4362",
    archivePrefix = "arXiv",
    primaryClass = "astro-ph",
    doi = "10.1051/0004-6361/200811029",
    journal = "Astron. Astrophys.",
    volume = "496",
    pages = "399--412",
    year = "2009"
}

@article{Carenza:2021alz,
    author = "Carenza, Pierluca and Evoli, Carmelo and Giannotti, Maurizio and Mirizzi, Alessandro and Montanino, Daniele",
    title = "{Turbulent axion-photon conversions in the Milky~Way}",
    eprint = "2104.13935",
    archivePrefix = "arXiv",
    primaryClass = "hep-ph",
    doi = "10.1103/PhysRevD.104.023003",
    journal = "Phys. Rev. D",
    volume = "104",
    number = "2",
    pages = "023003",
    year = "2021"
}

@article{Meyer:2014epa,
    author = "Meyer, Manuel and Montanino, Daniele and Conrad, Jan",
    title = "{On detecting oscillations of gamma rays into axion-like particles in turbulent and coherent magnetic fields}",
    eprint = "1406.5972",
    archivePrefix = "arXiv",
    primaryClass = "astro-ph.HE",
    doi = "10.1088/1475-7516/2014/09/003",
    journal = "JCAP",
    volume = "09",
    pages = "003",
    year = "2014"
}

@article{Jansson:2012rt,
    author = "Jansson, Ronnie and Farrar, Glennys R.",
    title = "{The Galactic Magnetic Field}",
    eprint = "1210.7820",
    archivePrefix = "arXiv",
    primaryClass = "astro-ph.GA",
    doi = "10.1088/2041-8205/761/1/L11",
    journal = "Astrophys. J. Lett.",
    volume = "761",
    pages = "L11",
    year = "2012"
}

@article{Planck:2016gdp,
    author = "Adam, R. and others",
    collaboration = "Planck",
    title = "{Planck intermediate results.}: {XLII. Large-scale Galactic magnetic fields}",
    eprint = "1601.00546",
    archivePrefix = "arXiv",
    primaryClass = "astro-ph.GA",
    doi = "10.1051/0004-6361/201528033",
    journal = "Astron. Astrophys.",
    volume = "596",
    pages = "A103",
    year = "2016",
    note = "Our $B_{\rm rms}=5~\mu{\rm G}$ is a factor of $\sqrt{2}$ larger than given in the Planck paper, which used an older version of the Hammurabi code that was missing a factor of 2 in the normalization of synchrotron radiation (\href{https://sourceforge.net/projects/hammurabicode/files/}{https://sourceforge.net/projects/hammurabicode/files/}). Thanks to Michael Unger for pointing this out and to Tess Jaffe for confirmation."
}

@ARTICLE{2024MNRAS.535.1944L,
       author = {{Livingston}, J.~D. and {McClure-Griffiths}, N.~M. and {Ma}, Y.~K. and {Bustard}, C. and {Mao}, S.~A. and {Gaensler}, B.~M. and {Kaczmarek}, J.},
        title = "{Magnetic fields in the Large Magellanic Cloud and their connection to the Magellanic System}",
      journal = {Mon. Not. R. Astron. Soc.},
         year = 2024,
        month = dec,
       volume = {535},
       number = {2},
        pages = {1944-1963},
          doi = {10.1093/mnras/stae2416},
       adsurl = {https://ui.adsabs.harvard.edu/abs/2024MNRAS.535.1944L}
}

@article{Haverkorn_2013,
   title={Plasma Diagnostics of the Interstellar Medium with Radio Astronomy},
   volume={178},
   ISSN={1572-9672},
   url={http://dx.doi.org/10.1007/s11214-013-0014-6},
   DOI={10.1007/s11214-013-0014-6},
   number={2–4},
   journal={Space Science Reviews},
   publisher={Springer Science and Business Media LLC},
   author={Haverkorn, Marijke and Spangler, Steven R.},
   year={2013},
   pages={483–511} }

@article{Mirizzi:2007hr,
    author = "Mirizzi, Alessandro and Raffelt, Georg G. and Serpico, Pasquale D.",
    title = "{Signatures of Axion-Like Particles in the Spectra of TeV Gamma-Ray Sources}",
    eprint = "0704.3044",
    archivePrefix = "arXiv",
    primaryClass = "astro-ph",
    reportNumber = "FERMILAB-PUB-07-082-A, MPP-2007-44",
    doi = "10.1103/PhysRevD.76.023001",
    journal = "Phys. Rev. D",
    volume = "76",
    pages = "023001",
    year = "2007"
}

@misc{SM,
 note = {Supplemental Material. It includes bounds on $g_{ap}\times g_{a\gamma}$ when varying the assumed CL, bounds on $g_{a\gamma}$ alone, a discussion on dissipation scale for the turbulence and its structure function. Finally, we detail how fluctuations in the magnetic field affect the conversion probability in the massless and heavy axion regimes. It includes \cite{CAST:2024eil,Noordhuis:2022ljw,kolmogorov1991local,padmanabhan2000theoretical,Haverkorn_2013}}
}
\bibliographystyle{bibi}

\onecolumngrid

\onecolumngrid
\appendix

\setcounter{equation}{0}
\setcounter{figure}{0}
\setcounter{table}{0}
\setcounter{page}{1}
\makeatletter
\renewcommand{\theequation}{S\arabic{equation}}
\renewcommand{\thefigure}{S\arabic{figure}}
\renewcommand{\thepage}{S\arabic{page}}

\begin{center}
\textbf{\large Supplemental Material for the Letter\\[0.5ex]
{\em Magnetic Turbulence Boosts Supernova Signals of Axion--Photon Conversion}}
\end{center}

In this Supplemental Material, we discuss how the constraints on $g_{ap}\times g_{a\gamma}$ obtained in the main text change when varying the assumed CL, and show the bounds on $g_{a\gamma}$ alone. Furthermore, we explicitly estimate the dissipation scale for the turbulence and recollect its structure function from the literature. Finally, we detail how fluctuations in the magnetic field affect the conversion probability in the massless and heavy axion regimes.

\bigskip

\begin{figure}[h!]
    \centering
    \includegraphics[width=0.5\linewidth]{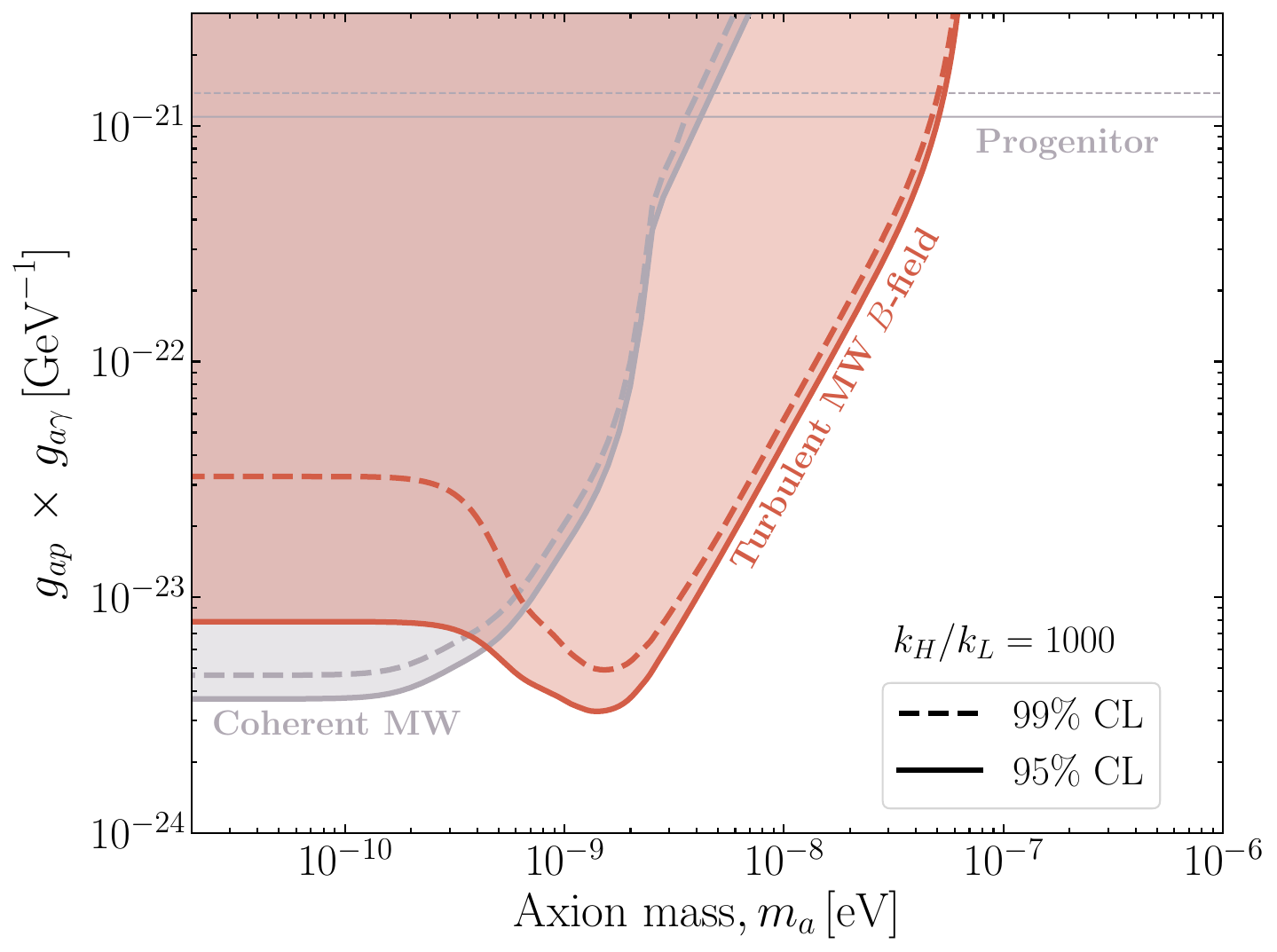}
    \caption{Like Fig.~\ref{fig:BremssBounds}, now only for the MW. We show the bounds for confidence levels 95\% and 99.73\%. The impact of the uncertain $B$-field configuration along the LoS for turbulent fields is only strong in the massless limit.}
    \label{fig:BoundsSuppMat}
\end{figure}

\textbf{\textit{\boldmath Dependence of bounds on the confidence level}}---In Fig.~\ref{fig:BremssBounds} of the main text, we showed bounds at 95\%~CL, which is a typical confidence level chosen for the statistical power of astrophysical bounds. For completeness, we now also show the $3\sigma$ bounds with a confidence level of $99.7\%$. This is shown in Fig.~\ref{fig:BoundsSuppMat} for the MW field, compared with the $95\%$~CL bounds, both for the regular and the turbulent field.

For the coherent $B$-field, the modification is uniform across all $m_a$ values: the constraint on a possible signal at SMM is $S<31.7$ for 99.73\%~CL and $S<20.1$ for 95\% CL, so the bounds change only by $30\%$. However, in the turbulent case, $S$ itself is a random variable whose dispersion depends on $m_a$. Since the specific turbulent configuration along our line of sight may considerably fluctuate, changing the confidence level has more of an impact in this case. This is especially true for the massless regime, where the conversion effectively probes only a single quantity, namely the magnetic field averaged over the line of sight, which therefore fluctuates significantly. Thus, we find $S<2706.9$ for 99.73\%~CL and $S<158.8$ for 95\% CL, i.e., the bound is weakened by a factor of 4. On the other hand, heavier axions---our main regime of interest---probe an entire range of wavenumbers around the resonant one, since they have an energy spectrum, and thus the impact of fluctuations is dramatically reduced. For example, for $m_a=3\, \mathrm{neV}$, we find $S<42.7$ at 99.73\%~CL and $S<24.1$ at 95\% CL---a factor 1.3 weaker.

\begin{figure}
    \centering
    \includegraphics[width=0.5\linewidth]{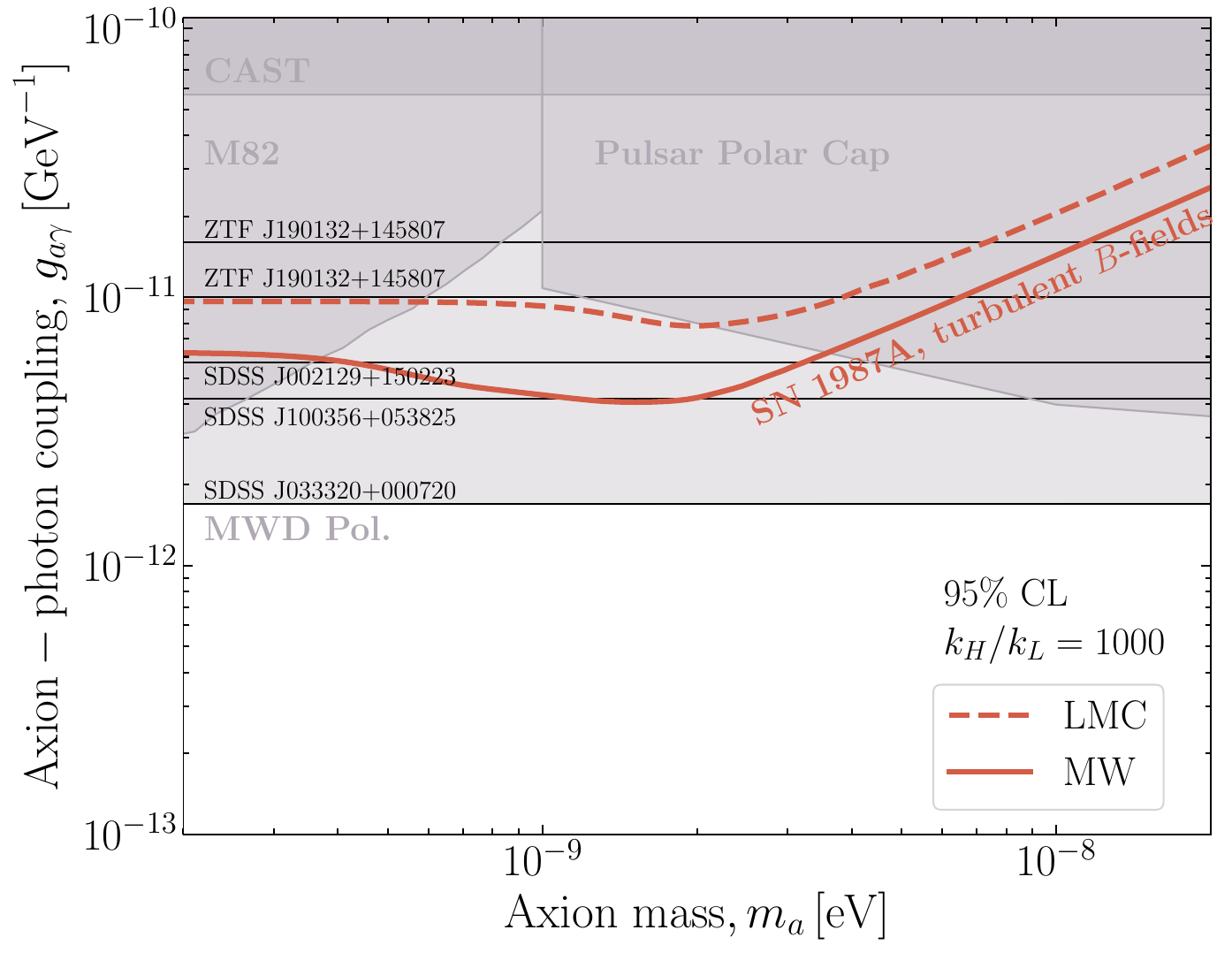}
    \caption{Like Fig.~\ref{fig:BremssBounds}, now for the axion-photon coupling $g_{a\gamma}$ alone, assuming production by Primakoff emission. Limits from other sources (grey shaded) are taken from the compilation of Ref.~\cite{OHare}; specifically, we show bounds from M82~\cite{Ning:2024eky}, CAST~\cite{CAST:2024eil}, Pulsar Polar Cap \cite{Noordhuis:2022ljw}, and different magnetic white dwarfs~\cite{Benabou:2025jcv} as indicated.}
    \label{fig:boundsPrimakoff}
\end{figure}

\textbf{\textit{\boldmath Bound on $g_{a\gamma}$ from Primakoff emission alone}}---In the main text, we assumed that axions are produced by $np$-bremsstrahlung due to the interaction with protons. However, we may also constrain $g_{a\gamma}$ alone by assuming production by Primakoff scattering alone. We assume the quasi-thermal spectrum of Eq.~\eqref{eq:GammaSpectrum} with the parameter values shown in Table~\ref{tab:Fluxes}. Notice that the Gamma distribution of the expected number of signal events at SMM is not the same as for the bremsstrahlung sourced flux, since the axion spectrum is different and so the values of the shape paramter $\kappa$ will differ. The results are shown in Fig.~\ref{fig:boundsPrimakoff}: for axion masses $3\times10^{-10}\, \mathrm{eV} \lesssim m_a\lesssim 3\times 10^{-9}\, \mathrm{eV}$, our bounds are among the most stringent ones, competing with existing ones from the lack of observation of linearly polarized light from magnetic white dwarfs~\cite{Benabou:2025jcv}.

\textbf{\textit{Dissipation scale for turbulence}}---A primary source of uncertainty in our treatment is the length scale at which the turbulent power spectrum is suppressed. As the resolution of current observations is insufficient to determine it, we turn to a simple theoretical estimate. The turbulent cascade is interrupted at scales such that damping is faster than the cascade timescale. Assuming for a moment Kolmogorov turbulence~\cite{kolmogorov1991local} (we discuss later how the results should change for the more realistic case of magnetized turbulence), the cascade timescale can be estimated as $t_{\rm cas}\sim \ell/u_\ell$, where $\ell$ is the length scale perpendicular to the local magnetic field and $u_\ell$ is the typical velocity fluctuation at that scale. In Kolmogorov turbulence, we can relate $u_\ell$ with the velocity $u_{\ell_{\rm out}}$ at the outer scale $ \ell_{\rm out}$ as
\begin{equation}
    u_\ell\sim u_{\ell_{\rm out}} \left(\frac{\ell}{\ell_{\rm out}}\right)^{1/3}.
\end{equation}
If the magnetic field is mostly turbulent, as measurements seem to suggest, at the outer scale the dynamo mechanism would likely induce equipartition of magnetic and kinetic energy, so we may take $u_{\ell_{\rm out}}\sim v_A$, the Alfv{\'e}n velocity. We now turn to the timescale for damping of the excitations. The primary damping mechanism in a partially ionized gas is neutral-ion damping: the force acting on both ion and neutral components is parameterized as
\begin{equation}
    \mathbf{F}=\gamma \rho_n \rho_i (\bv_n-\bv_i),
\end{equation}
where $\bv_i$ and $\bv_n$ are the bulk velocities of both components, $\rho_i$ and $\rho_n$ their densities, and $\gamma\sim 10^{14}\,\mathrm{cm}^3\,\mathrm{g}^{-1}\,\mathrm{s}^{-1}$~\cite{padmanabhan2000theoretical}. Balancing this force, acting on ions, with the Lorentz force $\mathbf{F}=\mathbf{j}\times \bB=(\boldsymbol{\nabla}\times \bB)\times \bB/4\pi$ (for easier comparison with the literature we use unrationalized units), we see that the typical velocity drift $\delta\bv_\ell=(\bv_n-\bv_i)_\ell$ at length scale $\ell$ is
\begin{equation}
    \delta v_\ell\sim \frac{B_0 \delta B_\ell}{4\pi \ell \gamma \rho_n \rho_i},
\end{equation}
where $B_0$ is the outer-scale field, which looks coherent from the smaller scales, while $\delta B_\ell$ is the fluctuating field at the length scale $\ell$. This estimate is conceptually valid only if ions and neutrals are strongly coupled on the frequencies of the magnetic perturbation, requiring $\gamma \rho_n \gg \omega_\ell\sim v_a/\ell_\parallel$, where $\ell_\parallel$ is the wavelength of the perturbation parallel to the field at the outer scale. With the numbers we use later, this inequality is usually satisfied.  The dissipated energy per unit time at length scale $\ell$ is then $Q_\ell=\mathbf{F}_\ell\cdot \delta \bv_\ell$, so that
\begin{equation}
    Q_\ell\sim \frac{B_0^2 \delta B_\ell^2}{16\pi^2 \ell^2 \gamma \rho_n \rho_i}\sim \frac{ v_A^2}{\ell^2\gamma \rho_n \rho_i} U_\ell,
\end{equation}
where $U_\ell\sim \delta B_\ell^2/4\pi$ is the fluctuation energy at length scale $\ell$ (we include a factor of 2 to account for the near-equipartitioned magnetic and kinetic energy), and we have used the definition of the Alfv{\'e}n velocity $v_A^2=B_0^2/4\pi \rho_{\rm tot}$---the total density $\rho_{\rm tot}=\rho_n+\rho_i$ appears because ions and neutrals are strongly coupled. Therefore, the damping timescale of turbulent fluctuations at length scale $\ell$ is
\begin{equation}\label{eq:damping_timescale}
    t_{\rm damp}\sim \frac{\gamma \rho_n  \rho_i \ell^2}{\rho_{\rm tot}v_A^2}.
\end{equation}
Balancing the damping and the cascade rate, we find that turbulent fluctuations should be suppressed below a length scale
\begin{equation}
    \ell_{\rm damp}\sim \ell_{\rm out} \left(\frac{v_A \rho_{\rm tot}}{\gamma \rho_n \rho_i \ell_{\rm out}}\right)^{3/4}.
\end{equation}
Since $\ell_{\rm out}\sim 100\,\mathrm{pc}$, assuming a mildly ionized medium with $\rho_n\sim \rho_i\sim 10^{-24}\,\mathrm{g}\,\mathrm{cm}^{-3}$, we find that the cascade should survive down to length scales $\ell_{\rm damp}/\ell_{\rm out}\sim 10^{-3}$, which is the ratio we used for our results.

On the other hand, this estimate is likely overly conservative. The assumption of Kolmogorov turbulence is inconsistent if the magnetic field at the injection scale is at equipartition. In this case, we should instead turn to the Goldreich-Sridhar (GS) picture~\cite{Goldreich:1994zz}, in which the cascade proceeds with the same Kolmogorov dependence in the length scale perpendicular to the local magnetic field $\ell_\perp$. However, the parallel length scale remains much larger, fixed by the condition of critical balance $u_\ell/\ell_\perp\sim v_A/\ell_\parallel$, so $\ell_\parallel\sim L(\ell_\perp/L)^{2/3}$. This significantly enhances $k_H$, since in the damping timescale in Eq.~\eqref{eq:damping_timescale} it is the parallel length scale driving dissipation; indeed, in the Lorentz force term $(\boldsymbol{\nabla}\times \bB)\times \bB$, only the parallel derivative of the transverse Alfv{\'e}nic field contributes. Therefore, the correct condition for the damping length scale is
\begin{equation}
    \frac{\gamma \rho_n \rho_i \ell_\parallel^2}{\rho_{\rm tot} v_A^2}\sim \frac{\ell_\perp}{u_\ell}.
\end{equation}
This leads to a much smaller damping length scale
\begin{equation}
    \ell_{\rm damp}\sim \ell_{\rm out} \left(\frac{v_A \rho_{\rm tot}}{\gamma \rho_n \rho_i \ell_{\rm out}}\right)^{3/2},
\end{equation}
so that $\ell_{\rm damp}/\ell_{\rm out}\sim 10^{-6}$. This is much easier to reconcile with the observation that the density fluctuations in the warm ionized interstellar medium follow an unbroken Kolmogorov spectrum for at least five decades~\cite{Armstrong:1995zc}; we refer to the detailed discussion in Ref.~\cite{Haverkorn_2013} and especially Sec.~6.6 therein.

\textbf{\textit{Structure functions of turbulence}}---Let us consider the general structure of turbulence, which we take to be isotropic. The core structure function is taken to be
\begin{equation}
    \langle B_{i,\bk}B_{j,\bk'}^*\rangle=\left(\delta_{ij}-\frac{k_i k_j}{k^2}\right)(2\pi)^3 \delta(\bk-\bk')\frac{\pi^2 (\beta-1) B_0^2}{k_L^{-\beta+1}-k_H^{-\beta+1}}k^{-\beta-2},
\end{equation}
where $k=|\bk|$ and the structure function vanishes outside the range $k_L<k<k_H$. The normalization is already chosen such that $\langle |\bB(\br)|^2\rangle=B_0^2$.

Let us now consider the structure function of turbulence along a specific line of sight, which we take to be along the $z$ axis. Thus we seek
\begin{equation}
    \langle B_x(z_1) B_x(z_2)\rangle=\langle B_y(z_1) B_y(z_2)\rangle=\int_{-\infty}^{+\infty}\frac{dk_z}{2\pi}e^{ik_z(z_1-z_2)}\int \frac{d^2\bq}{(2\pi)^2}\left[1-\frac{q^2 \cos^2\phi}{q^2+k_z^2}\right]\frac{\pi^2 (\beta-1) B_0^2}{k_L^{-\beta+1}-k_H^{-\beta+1}}(k_z^2+q^2)^{-\frac{\beta}{2}-1},
\end{equation}
where $\bq=(q\cos\phi, q\sin \phi)=(k_x,k_y)$. We can write this in the form
\begin{equation}
    \langle B_x(z_1) B_x(z_2)\rangle =B_0^2\int_0^{k_H} dk_z \Psi_{k_z}\cos\left[k_z(z_1-z_2)\right],
\end{equation}
to match with the definitions in the next section. The function $\Psi_{k_z}$ then has the form
\begin{equation}
    \Psi_{k_z}=\frac{(\beta-1)}{k_L^{-\beta+1}-k_H^{-\beta+1}}\int \frac{q dq}{2}\left[1-\frac{q^2}{2(q^2+k_z^2)}\right](k_z^2+q^2)^{-\frac{\beta}{2}-1},
\end{equation}
where the integral is done only over $k_L^2<q^2+k_z^2<k_H^2$. For $k_L\ll k_H$, we can neglect in the denominator the term $k_H^{-\beta+1}$. The integral over $q$ can be done analytically, giving finally
\begin{equation}
    \Psi_{k_z}=\frac{(\beta-1)k_L^{\beta-1}}{4}\left[\frac{k_{\rm min}^{-\beta}-k_H^{-\beta}}{\beta}+\frac{k_z^2}{\beta+2}\left(k_{\rm min}^{-\beta-2}-k_H^{-\beta-2}\right)\right],
\end{equation}
where $k_{\rm min}=\mathrm{max}(|k_z|,k_L)$. The structure function for $\langle B_y(z_1) B_y(z_2)\rangle$ is of course analogous, while by the same manipulation we find that
\begin{equation}
    \langle B_z(z_1) B_z(z_2)\rangle=B_0^2 \int_0^{k_H}dk_z \Gamma_{k_z}\cos\left[k_z(z_1-z_2)\right],
\end{equation}
where
\begin{equation}
    \Gamma_{k_z}=\frac{(\beta-1)k_L^{\beta-1}}{2}\left[\frac{k_{\rm min}^{-\beta}-k_H^{-\beta}}{\beta}-\frac{k_z^2}{\beta+2}\left(k_{\rm min}^{-\beta-2}-k_H^{-\beta-2}\right)\right].
\end{equation}
Therefore, we finally find
\begin{equation}
    \langle \bB(z_1)\cdot \bB(z_2)\rangle=B_0^2 \int_0^{k_H}dk_z \Phi_{k_z}\cos\left[k_z(z_1-z_2)\right],
\end{equation}
with
\begin{equation}
    \Phi_{k_z}=(\beta-1) k_L^{\beta-1}\frac{k_{\rm min}^{-\beta}-k_H^{-\beta}}{\beta}.
\end{equation}
For axion-photon conversions, only the function $\Psi_{k_z}$ enters, since the longitudinal component cannot participate.

If we denote the Fourier transform of the turbulence by
\begin{equation}
    B_{x,k_z}=\int B_x(z) e^{-i k_z (z-L/2)}dz,
\end{equation}
the probability distribution of the Fourier coefficients must therefore be
\begin{equation}
    P\left\{B_{x,k_z}\right\}\propto \exp\left[-\int_0^{k_H} \frac{dk_z}{2\pi}\frac{|B_{x,k_z}|^2}{\pi B_0^2 \Psi_{k_z}}\right].
\end{equation}
For numerical purposes, it is convenient to expand the magnetic field as a periodic function in a Fourier series with period $L$, rather than in continuous Fourier components. For the conversion probability, this makes of course no difference. We may therefore express
\begin{equation}
    B_{x}(z)=\sum_{n=-\infty}^{+\infty}b_{x,n} e^{ik_n (z-L/2)},
\end{equation}
with $k_n=2\pi n/L$.
The variables $b_{x,n}$ are also normally distributed, and their variance can be found by the inverse Fourier coefficients
\begin{equation}
    \langle |b_{x,n}|^2\rangle=\frac{B_0^2}{2L^2}\int_0^{k_H}dk_z K^2(k_z-k_n)\Psi_{k_z}\simeq \frac{\pi B_0^2}{L}\Psi_{k_n},
\end{equation}
where $K(x)=2\sin(xL/2)/x$, and we have used the approximate equality $K^2(x)\to 2\pi L \delta(x)$ since the spectrum $\Psi_k$ changes over ranges of wavenumbers much larger than $2\pi/L$. The coefficients can be split into their real and imaginary parts $\langle b^{R,2}_{x,n}\rangle=\langle b^{I,2}_{x,n}\rangle=\langle |b_{x,n}|^2\rangle/2$, except for $n=0$ which is purely real so $\langle b_{x,0}^2\rangle=\pi B_0^2 \Psi_{0}/L$.

The conversion amplitude is now
\begin{equation}
    \mathcal{A}_x(k_a)=\frac{g_{a\gamma}}{2}\int_0^L dz \sum_{n=-\infty}^{+\infty} b_{x,n} e^{i(k_n-k_a)(z-L/2)}=\frac{g_{a\gamma}}{2}\left\{b_{x,0} K(k_a)+\sum_{n=1}^{+\infty}\left[b_{x,n} K(k_n-k_a)+b_{x,n}^* K(k_n+k_a)\right]\right\}.
\end{equation}
The conversion probability is $P_{a\gamma}(k_a)=|\mathcal{A}_x|^2+|\mathcal{A}_y|^2$. The variables $b_{x,n}=b^R_{x,n}+ib^I_{x,n}$ are sampled from a Gaussian distribution with $\langle b^{R,2}_{x,n}\rangle=\langle b^{I,2}_{x,n}\rangle=\pi B_0^2 \Psi(k_n)/2L$. Notice that the discrete Fourier coefficients are related to the continuous Fourier transform simply by $b_{x,n}=B_x(k_n)/L$. Thus, the expected conversion probability is
\begin{equation}
    \langle P_{a\gamma}(k_a)\rangle=\langle |\mathcal{A}_x|^2+|\mathcal{A}_y|^2\rangle=g_{a\gamma}^2\frac{\pi B_0^2}{2L}\left[\sum_{n=1}^{+\infty}\Psi_{k_n}\left[K^2(k_n+k_a)+K^2(k_n-k_a)\right]+\Psi_{k_0}K^2(k_a)\right].
\end{equation}
In the resonant approximation, this becomes
\begin{equation}\label{eq:prob_resonant}
    \langle P_{a\gamma}(k_a)\rangle=\frac{\pi}{2} g_{a\gamma}^2 B_0^2 L \Psi_{k_a},
\end{equation}
which continues smoothly also for $k_a\to 0$. This result is of course completely identical to what we would have obtained using the continuous Fourier transform: in that case
\begin{equation}
    \mathcal{A}_x(k_a)=\frac{g_{a\gamma}}{2}\int_0^{\infty} \frac{dk_z}{2\pi} \left[B_{x,k_z} K(k_z-k_a)+B_{x,k_z}^* K(k_z+k_a)\right],
\end{equation}
and we easily see that $\langle P_{a\gamma}(k_a)\rangle$ coincides with the one obtained above.

\textbf{\textit{Fluctuations in the conversion probability}}---So far, we have considered only the energy-dependent average conversion probability. To move further, and consider the fluctuations, we first have to introduce the spectral averaging of the conversion probability over a distribution of energies, or, equivalently, of $k_a$
\begin{equation}
    P_x=\int_0^{\infty}dk_a \mathcal{S}(k_a)|\mathcal{A}_x(k_a)|^2,
\end{equation}
and similarly for $P_y$. The distribution $\mathcal{S}(k_a)$ is assumed normalized to unity. The average value of $P_x$ is easily obtained
\begin{equation}
    \langle P_x\rangle=\frac{g_{a\gamma}^2 B_0^2}{4}\int_0^{\infty}dk_a \mathcal{S}(k_a)\int_0^{\infty}dk_z \Psi_{k_z}\frac{K^2(k_z-k_a)+K^2(k_z+k_a)}{2};
\end{equation}
with the usual delta replacement, this becomes $\langle P_x\rangle =g_{a\gamma}^2 B_0^2 \pi L\int dk_a \mathcal{S}(k_a)\Psi_{k_a}/4$, as expected from Eq.~\eqref{eq:prob_resonant}.

We can now turn to the fluctuations. Here an exact expression, which is found by directly decomposing the quartic structure functions into all the possible quadratic combinations, is
\begin{align}
    \langle P_x^2\rangle -\langle P_x\rangle^2&=\frac{g_{a\gamma}^4 \pi^2 B_0^4}{16}\int dk_a \mathcal{S}(k_a)\int dk_b \mathcal{S}(k_b)\int\frac{dk_1}{2\pi}\int \frac{dk_2}{2\pi}\Psi_{k_1}\Psi_{k_2}
    \nonumber
    \\
    &{\kern2em }\times\kern-0.5em\sum_{\sigma_1,\sigma_2,\sigma_r=-1}^{+1}\left[K(k_1-\sigma_1k_a)K(k_1-\sigma_1 \sigma_r k_b)K(k_2-\sigma_2 k_a)K(k_2-\sigma_2 \sigma_r k_b)\right].
\end{align}
While this form is not particularly illuminating, we can now treat separately the regimes $k_a\ll 2\pi/L$ and $k_a\gg 2\pi/L$. In the first regime, $k_a$ can be neglected altogether, so we obtain
\begin{equation}
    \langle P_x\rangle=\frac{\pi}{4} g_{a\gamma}^2 B_0^2 L \Psi_{k_z=0}
    \quad\text{and}\quad
    \langle P_x^2\rangle -\langle P_x\rangle^2= 2\langle P_x\rangle^2.
\end{equation}
Since $P_{a\gamma}=P_x+P_y$, it follows that $\langle P^2\rangle-\langle P\rangle^2=\langle P\rangle^2$, i.e., $P$ has a variance equal to its expected value. In fact, its distribution is just exponential, since in this limit $\mathcal{A}_x(k_a=0)$ is an integral of Gaussian variables, and therefore its square $P_x$ is distributed according to a chi squared distribution with one degree of freedom. Therefore, the sum of $P_x$ and $P_y$ is distributed as a chi squared with two degrees of freedom, i.e., an exponential distribution.

Let us turn to the range $2\pi/L\ll k_a\ll k_H$. In this case, the main simplification is to only consider the contributions from $K(k_{1,2}-k_{a,b})$, which are the only ones benefiting from the resonant enhancement. We can use the usual result $K^2(x)\to 2\pi L \delta(x)$, and also that
\begin{equation}
    \int_{-\infty}^{+\infty}\frac{dk_1}{2\pi}K(k_1-k_a)K(k_1-k_b)=K(k_a-k_b)
\end{equation}
since we can extend the integration to negative $k_1$ as well where there is negligible contribution, and so we find
\begin{equation}
    \langle P_x\rangle=\frac{\pi}{4} g_{a\gamma}^2 B_0^2 L \int_0^{\infty} dk_a \mathcal{S}(k_a)\Psi_{k_a}
    \quad\text{and}\quad
    \langle P_x^2\rangle -\langle P_x\rangle^2=\frac{\pi^3}{8} g_{a\gamma}^4 B_0^4L \int_0^{\infty}dk_a \mathcal{S}^2(k_a)\Psi^2_{k_a}.
\end{equation}
The variance and mean of the total probability is simply twice as large, from the contribution due to $P_y$. Notice that the variance only grows as $L$, so in the limit of large $L$, the relative uncertainty drops as $L^{-1/2}$. Finally, in the very massive regime $k_a\gg k_H$, we may neglect $k_1$ and $k_2$ in the integrals, finally obtaining
\begin{equation}
    \langle P_x\rangle =\frac{g_{a\gamma}^2 \langle B_x(z)^2\rangle}{4} \int dk_a \mathcal{S}(k_a) K^2(k_a)\simeq \frac{g_{a\gamma}^2 \langle B_x(z)\rangle^2}{2}\int dk_a \frac{\mathcal{S}(k_a)}{k_a^2}
    \quad\text{and}\quad
    \langle P_x^2\rangle -\langle P_x\rangle^2=2\langle P_x\rangle^2.
\end{equation}

\begin{figure*}
\centering
\subfloat{
  \centering
  \includegraphics[width=0.48\linewidth]{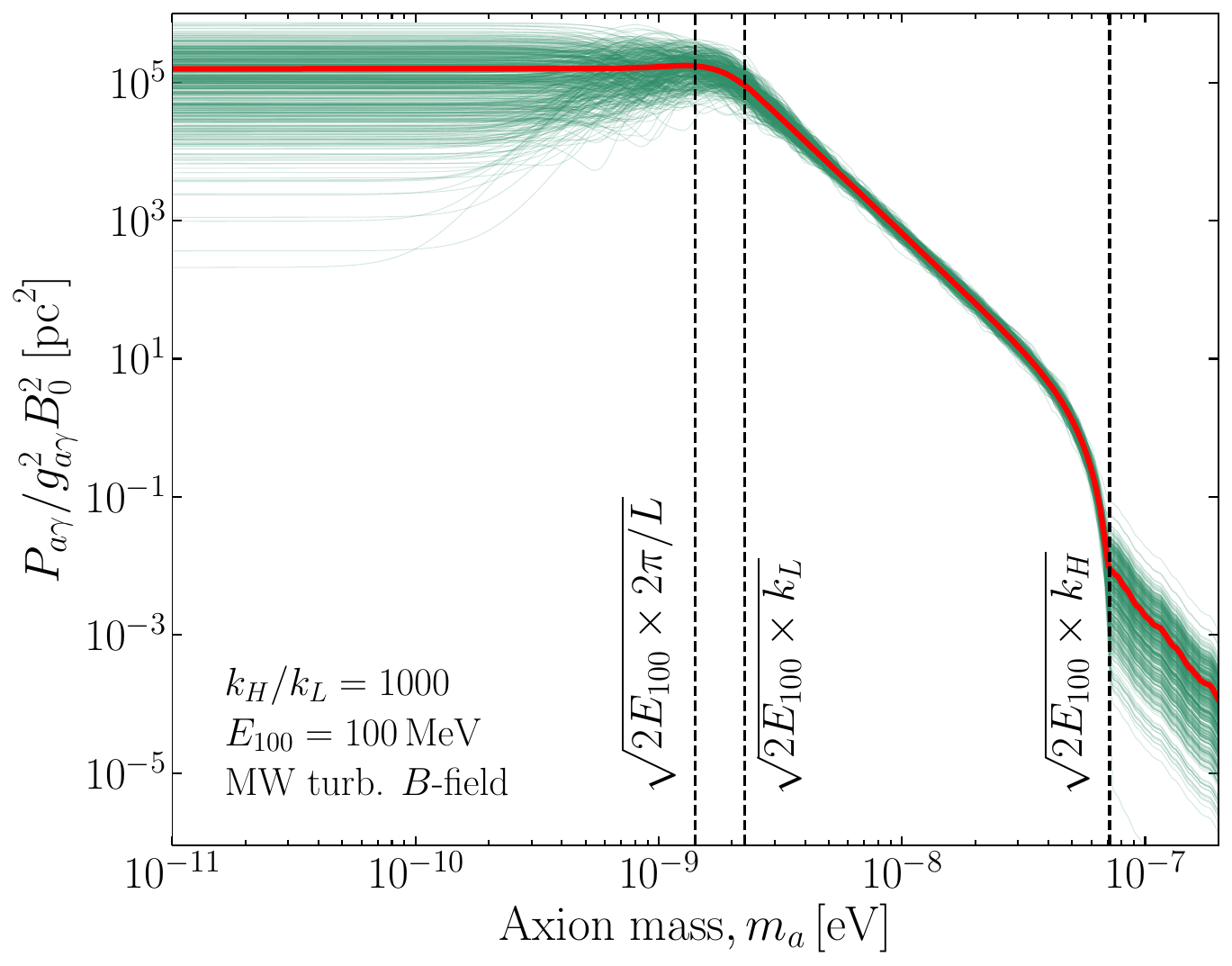}}
\subfloat{
  \centering
  \includegraphics[width=0.48\linewidth]{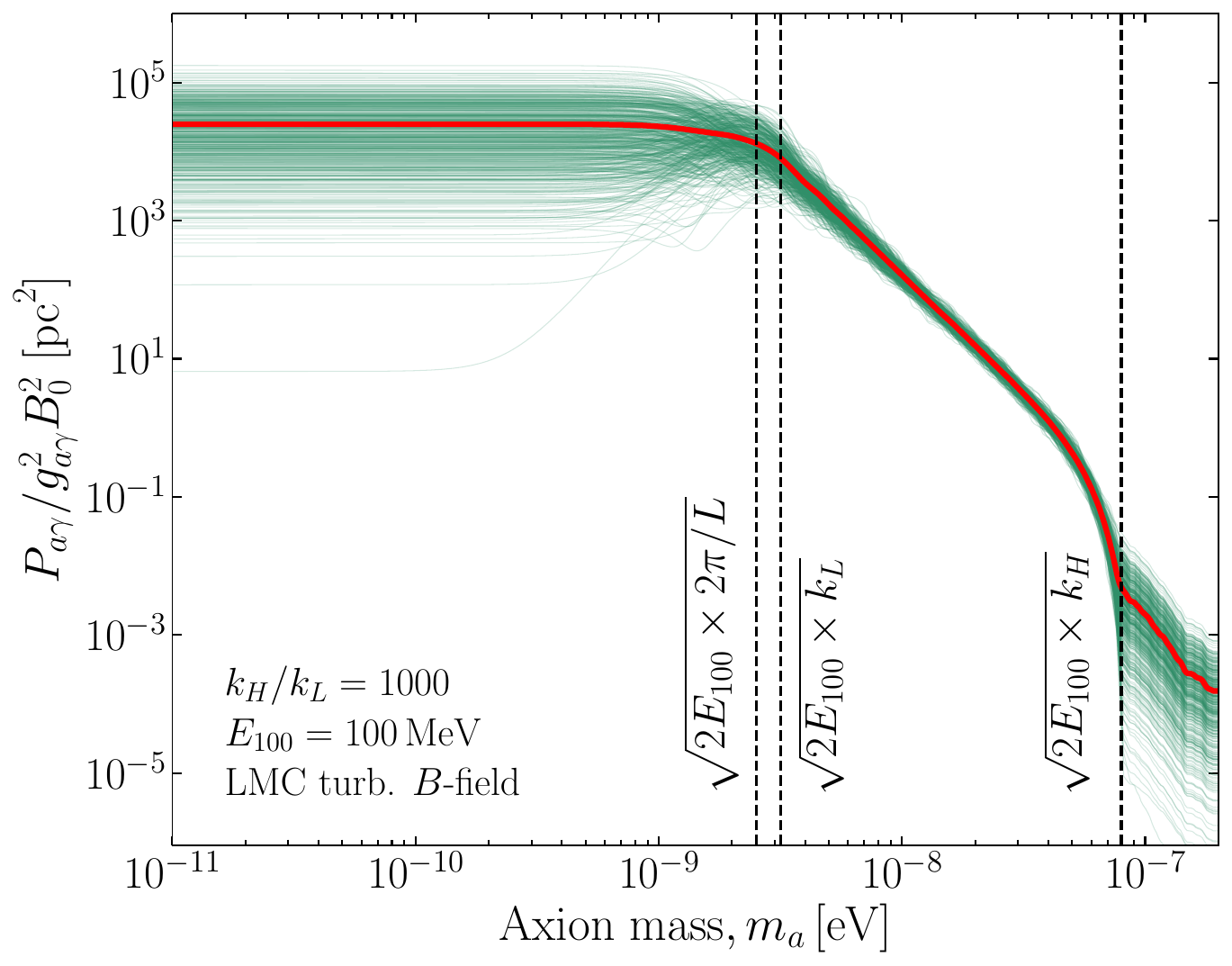}}
\caption{Same as Fig.~\ref{fig:mc} of the main text, but integrating the bremsstrahlung axion spectrum only in the 25--$100\, \mathrm{MeV}$ energy range observedby SMM, instead of the whole energy spectrum, for the MW (\textit{left panel}) and LMC (\textit{right panel}) turbulent $B$-fields. We use the values reported in Table~\ref{tab:Bfields} for the strength of the $B$-fields and in Table~\ref{tab:Fluxes} for the bremsstrahlung emission.}
\label{fig:mcSupM}
\end{figure*}

To verify these features, we perform a Monte Carlo simulation of the axion-photon conversion, generating $10^4$ realizations of synthetic turbulence realizations according to the Gaussian distribution. Our simulation is performed separately for the turbulent field in the Milky Way (MW) and in the Large Magellanic Cloud (LMC), which have different values of $k_H$, $k_L$, and $L$, as collected in Table~\ref{tab:Bfields}. In Fig.~\ref{fig:mcSupM}, we show 500 representative curves of conversion probability $P_{a\gamma}/g_{a\gamma}^2 B_0^2$, already averaged over the axion spectrum from bremsstrahlung emission in the 25--$100\, \mathrm{MeV}$ SMM window (in Fig.~\ref{fig:mc} of the main text we showed the conversion probability averaged over the whole spectrum).

Our analytical estimates are validated by the numerical results. At low axion masses, the resonant structures in turbulence have wavenumbers $k_z\lesssim k_L$, where the power spectrum is flat and the average conversion probability is independent from the mass. However, we identify two different regimes for the fluctuations: if the typical axion momentum $k_a\lesssim 2\pi/L$, the fluctuations around the average probability are very large. This is the regime where the probability is exponentially distributed, with a large number of realizations having a significantly lower conversion rate. For such low axion wavenumbers, axions are effectively resonating only with the zero-wavenumber mode, and there is a large chance that this might be underpopulated. When the axion momentum exceeds $k_a\gtrsim 2\pi/L$, the resonant target spans multiple wavenumbers---it is unlikely that all of them are underpopulated. Therefore, the fluctuations reduce considerably in amplitude. Above $k_a\gtrsim k_L$, the dominant axion target are turbulent fluctuations in the inertial range, and the conversion probability is suppressed by a power-law of the mass. Finally, at $k_a\gtrsim k_H$, no resonant target for conversion exists, so that there is a sharp drop in the conversion probability.

To provide a comprehensive description, it is useful to model the probability distribution for $P_{a\gamma}$ as a Gamma distribution
\begin{equation}\label{eq:GammaDistr}
    \mathcal{P}(P_{a\gamma})\propto P_{a\gamma}^{\kappa -1}\exp\left[-\frac{\kappa P_{a\gamma}}{\langle P_{a\gamma}\rangle}\right],
\end{equation}
where $\kappa$ is frequently called the shape parameter.
(Notice that the quasi-thermal spectrum of
Eq.~\eqref{eq:GammaSpectrum} is also a Gamma distribution, where traditionally the shape parameter or pinching parameter was chosen as $\alpha=\kappa-1$. Indeed, the larger $\alpha$, the smaller is the variance relative to the average, meaning the spectrum is more ``pinched.'') Matching both $\langle P_{a\gamma}\rangle$ and $\langle P_{a\gamma}^2\rangle$ to this distribution, we find that in the regime $k_a\ll 2\pi/L$ the distribution is exponential with
\begin{equation}
    \kappa =1
    \quad\text{and}\quad
    \langle P_{a\gamma}\rangle=\frac{\pi}{2} g_{a\gamma}^2 B_0^2 L \Psi_{k_z=0}.
\end{equation}
In the regime $2\pi/L\ll k_a\ll k_H$ instead we find
\begin{equation}\label{eq:intermkappa}
    \kappa =\frac{L}{\pi}\frac{\left[\int_0^{\infty}dk_a \mathcal{S}(k_a)\Psi_{k_a}\right]^2}{\int_0^{\infty}dk_a \mathcal{S}^2(k_a)\Psi^2_{k_a}}
    \quad\text{and}\quad
    \langle P_{a\gamma}\rangle=\frac{\pi}{2} g_{a\gamma}^2 B_0^2 L \int_0^{\infty}dk_a \mathcal{S}(k_a)\Psi_{k_a},
\end{equation}
and finally for $k_a\gg k_H$ again an exponential distribution with
\begin{equation}
    \kappa =1
    \quad\text{and}\quad
    \langle P_{a\gamma}\rangle=g_{a\gamma}^2 \langle B_x(z)\rangle^2\int_0^{\infty}dk_a\frac{\mathcal{S}(k_a)}{k_a^2}.
\end{equation}
The parameter $\kappa$ is actually the inverse of the coefficient of variation squared---exponential distributions are characterized by a coefficient of variation equal to 1, and so they largely fluctuate. This is the regime we find when $k_a\ll 2\pi/L$ and $k_a\gg k_H$. In the intermediate regime, the dispersion is much smaller, which leads to a larger value of $\kappa$ and a deviation from the exponential distribution. Indeed, for large $\kappa$, the Gamma distribution approaches a Gaussian one with ever smaller variance relative to its average.

\begin{figure*}
\centering
\subfloat{
  \centering
  \includegraphics[width=0.49\linewidth]{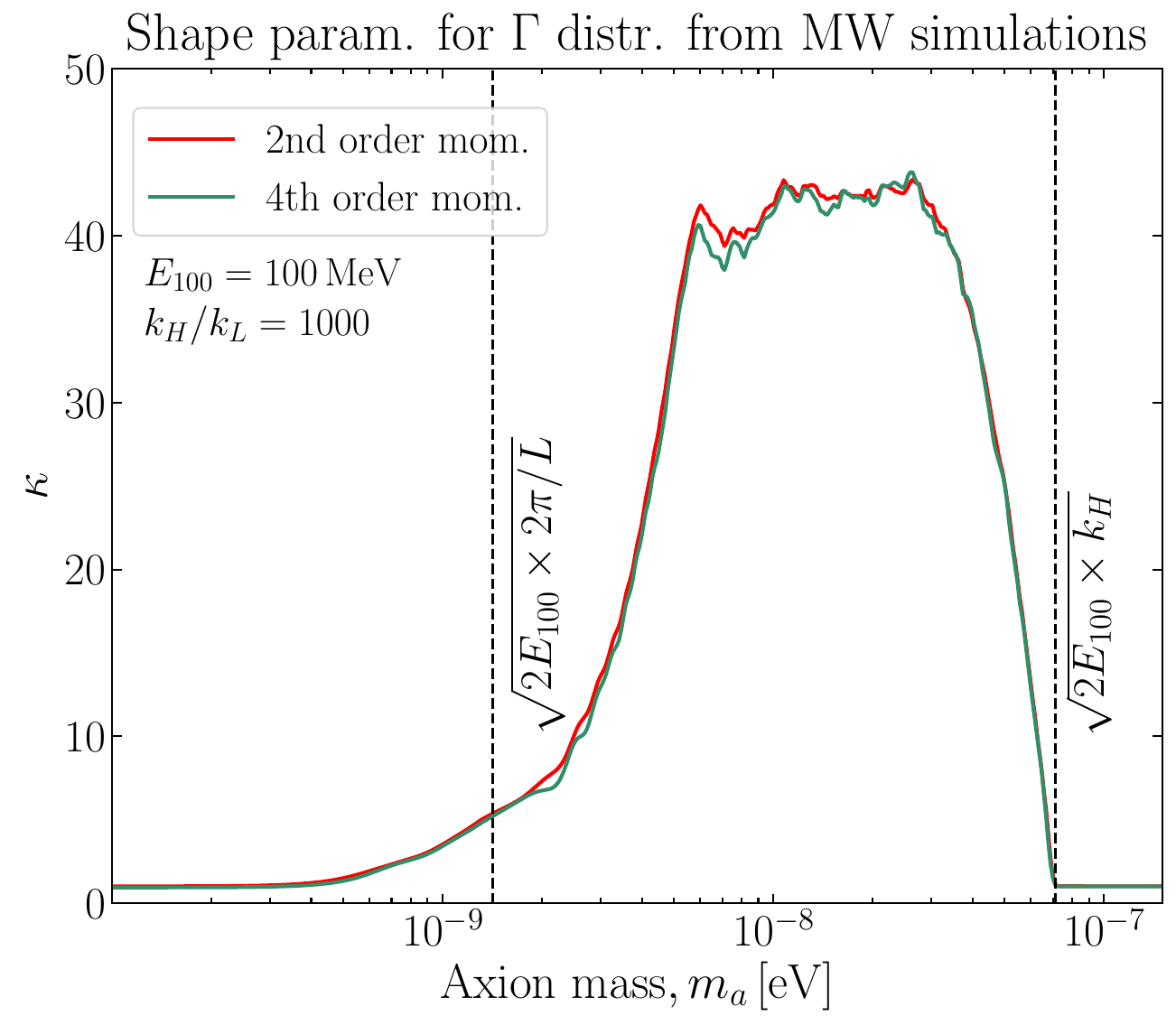}}
\subfloat{
  \centering
  \includegraphics[width=0.49\linewidth]{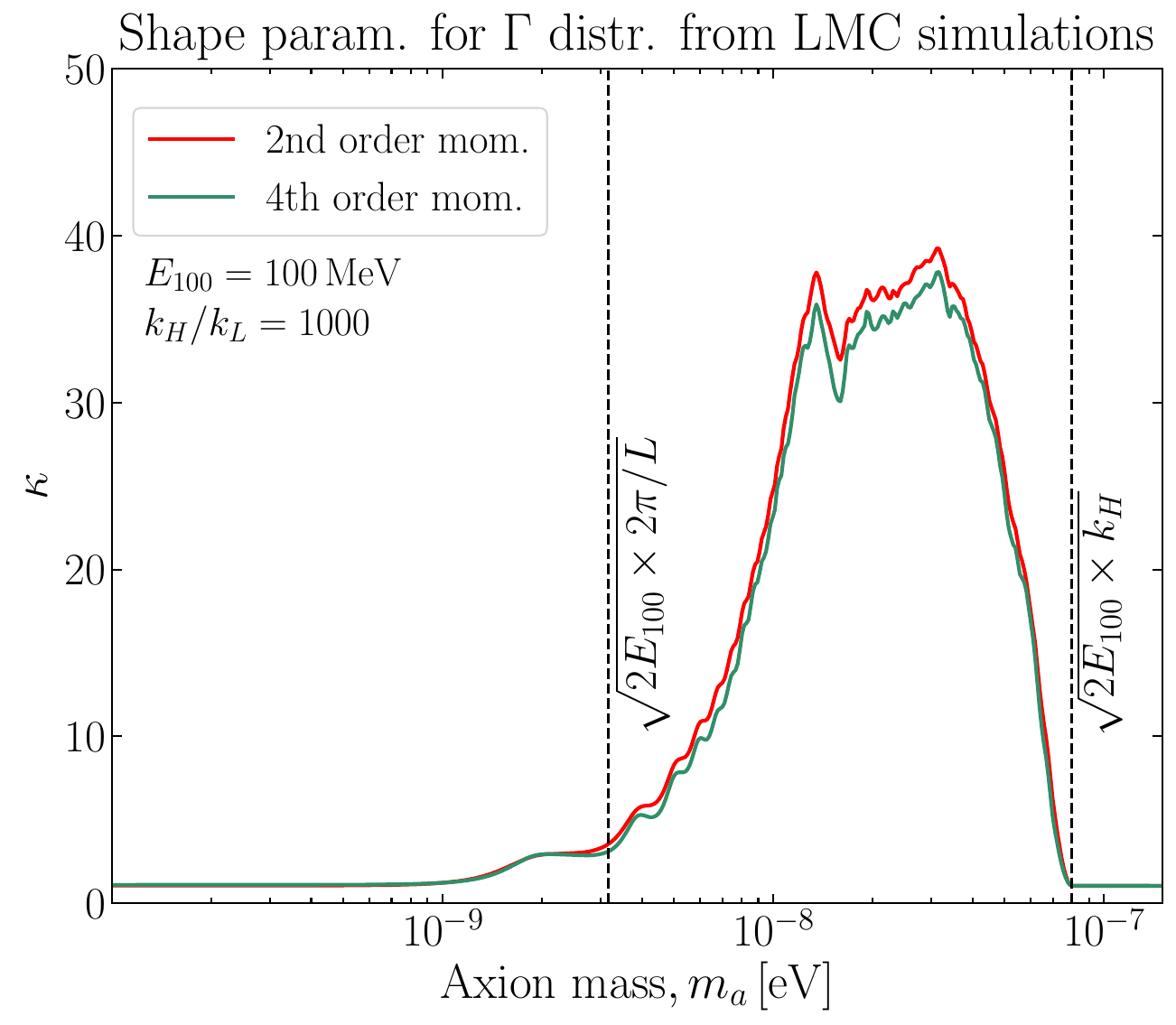}}
\caption{Fits for the shape parameter $\kappa$ for the Gamma distribution in Eq.~\eqref{eq:GammaDistr} using the second (red) and fourth (green) central moments from our $10^4$ Monte Carlo realizations of the turbulent $B$-field in MW (\textit{left panel}) and LMC (\textit{right panel}), convolved with the normalized axion spectrum in the 25--$100\, \mathrm{MeV}$ SMM window. We use the parameters of Table~\ref{tab:Fluxes} for the bremsstrahlung production. For all cases, $\kappa$ tends to 1 for $k\ll \sqrt{2 E_{100}\times 2\pi/L}$ and $k\gg \sqrt{2 E_{100}\times k_H}$, with $E_{100}=100\, \mathrm{MeV}$.}
\label{fig:kappas}
\end{figure*}

To derive the results shown in the main text, we obtained the value of $\kappa$ from our Monte Carlo simulations, using first (mean) and second (variance) central moments of our data. We confirmed that the Gamma distributions correctly describe our data by comparing the value of $\kappa$ obtained using the fourth central moment, related to the kurtosis, with one obtained using the variance (odd moments can be neglected, since they only carry information about the skewness). This comparison is shown in Fig.~\ref{fig:kappas}; the agreement on the estimate of the shape parameter is excellent. 

\end{document}